\documentclass[twocolumn]{aastex6}

\usepackage{graphicx}% Include figure files
\usepackage{epstopdf}
\usepackage{dcolumn}% Align table columns on decimal point
\usepackage{bm}% bold math
\usepackage{color}
\usepackage{epsfig}
\usepackage{amsmath}
\usepackage[caption=false]{subfig}
\def\ee{\end{equation}}
\def\be{\begin{equation}}
\def\l{\left}
\def\r{\right}
\def\pa{\partial}

\begin{document}


\title{Revisit of non-linear Landau damping for electrostatic instability driven by blazar-induced pair beams}

%% Use \author, \affil, plus the \and command to format author and affiliation
%% information.  If done correctly the peer review system will be able to
%% automatically put the author and affiliation information from the manuscript
%% and save the corresponding author the trouble of entering it by hand.
%%
%% The \affil should be used to document primary affiliations and the
%% \altaffil should be used for secondary affiliations, titles, or email.

%% Authors with the same affiliation can be grouped in a single

\author{S. Vafin$^1$, P. J. Deka$^1$, M. Pohl$^{1,2}$, and A. Bohdan$^2$ }
\affil{1 Institute for Physics and Astronomy, University of Potsdam, D-14476 Potsdam, Germany\\
2 DESY, Platanenallee 6, D-15738 Zeuthen, Germany}

\email{deka@uni-potsdam.de}

\begin{abstract}
We revisit the effect of non-linear Landau (NL) damping on the electrostatic instability of blazar-induced pair beams, using a realistic pair-beam distribution. We employ a simplified 2D model in ${\bf k}$-space to study the evolution of the electric-field spectrum and to calculate the relaxation time of the beam. We demonstrate that the 2D model is an adequate representation of the 3D physics. We find that non-linear Landau damping, once it operates efficiently, transports essentially the entire wave energy to small wavenumbers where wave driving is weak or absent. The relaxation time also strongly depends on the IGM temperature, $T_\mathrm{IGM}$, and for $T_\mathrm{IGM}\ll10$ eV, and in the absence of any other damping mechanism, the relaxation time of the pair beam is longer than the inverse Compton (IC) scattering time. The weak late-time beam energy losses arise from the accumulation of wave energy at small $k$, that non-linearly drains the wave energy at the resonant $\mathbf{k}$ of the pair-beam instability. Any other dissipation process operating at small $k$ would reduce that wave-energy drain and hence lead to stronger pair-beam energy losses. As an example, collisions reduce the relaxation time by an order of magnitude, although their rate is very small. Other non-linear processes, such as the modulation instability, could provide additional damping of the non-resonant waves and dramatically reduce the relaxation time of the pair beam.  An accurate description of the spectral evolution of the electrostatic waves is crucial for calculating the relaxation time of the pair beam. 
\end{abstract}


\section{Introduction}\label{Intro}
The propagation of very high energy gamma-radiation ($E_\gamma>100$ GeV) and its reaction on the inter-galactic medium (IGM) have been actively studied in recent years both observationally \citep{Neronov09,Neronov10,Ackermann12,Frank13,Naurois15,Funk15} and theoretically \citep{Elyiv09,Taylor11,Broderick12,Puchwein12,RS12a,RS12,Miniati13,Sironi14,Broderick16,Rafighi17,Vafin18}. An important feature of high-energy photons is the ability to interact with the extra-galactic background light (EBL) producing ultra-relativistic electron-positron beams \citep{Gould66}. Due to the inverse Compton (IC) scattering, these pairs will
emit secondary photons \citep{1994ApJ...423L...5A} that can be also energetic enough to efficiently interact with the EBL. The observational analysis indicates indicates, however, that the measured gamma-ray signal in the GeV energy band is smaller than the predicted cascade emission assuming that the pairs lose their energy only due to the IC scattering \citep{Neronov10, Tavecchio10, Tavecchio11}. Thus, some other dissipation processes must be in play. 

One possible explanation relies on the existence of a large-scale hypothetical magnetic field in cosmic voids \citep{Elyiv09, Neronov10, Taylor11} which remains questionable, as no direct magnetic field observations are yet available, but would, if true, havle profound consequences for magnetogenesis in the universe. The predicted magnetic field should result in GeV halos (about TeV sources) that are, however, absent as was shown recently \citep{Broderick18}. 

An alternative model uses only the fact that an electrositron beam propagating through the IGM plasma is subject to the electrostatic (two-stream) instability resulting in beam energy dissipation into the plasma waves \citep{Broderick12,RS12,Vafin18,Shalaby18}. Whether or not that is faster than IC cooling depends on the ratio of the beam-relaxation time and the IC scattering time. The former is determined by the instability growth rate as well as non-linear processes such as the non-linear Landau (NL) damping \citep{Breizman72,Breizman90}  and the modulation instability \citep{Zakharov72,Papadopoulos75,Baikov77}. Using a simplified model, \citet{Miniati13} claimed that the NL damping can stabilize the realistic blazar-induced pair beam on the IC timescale making the effect of plasma instabilities negligible. Later, \citet{Chang14} rigorously solving the kinetic wave equation came to the opposite conclusion. Their result relies on the fact that there is a region at large wave numbers where the electrostatic instability dominates over the NL as well as LL (linear Landau) damping. In that region, the electrostatic instability can efficiently reduce the beam energy. However, the work by \citet{Chang14} seems to overestimate the growth rate assuming it to be constant at wave numbers $|{\bf k}|\geq \omega_{p,e}/c$ ($\omega_{p,e}=\sqrt{4\pi n_e e^2/m_e}$ is the electron plasma frequency, $n_e$ the electron number density, and $c$ the speed of light). This fact contradicts the realistic growth rate which rapidly decreases with the wave number \citep{Miniati13,Vafin18}. Therefore, the primary goal of this paper is to investigate the effect of the NL damping taking into account the realistic behavior of the electrostatic growth rate at large wave numbers for blazar-induced pair beams.

Other dissipation processes, such as the modulation instability and collisional damping, can also have a dramatic effect on the relaxation time of the pair beams. In the current work, we focus on the collisional damping in addition to the NL damping, whereas the modulation instability requires a more careful study going beyond the scope of the current manuscript.

The rest of the paper is organized as follows. Section \ref{theory} explains our theoretical model. Simulation results without and with the collisional damping are presented in section \ref{results} and \ref{collisions}, respectively.  The final summary of the work is given in section \ref{summary}.

\section{Theoretical model}\label{theory}

\subsection{Electrostatic growth rate}

\begin{figure}
\includegraphics[width = \columnwidth]{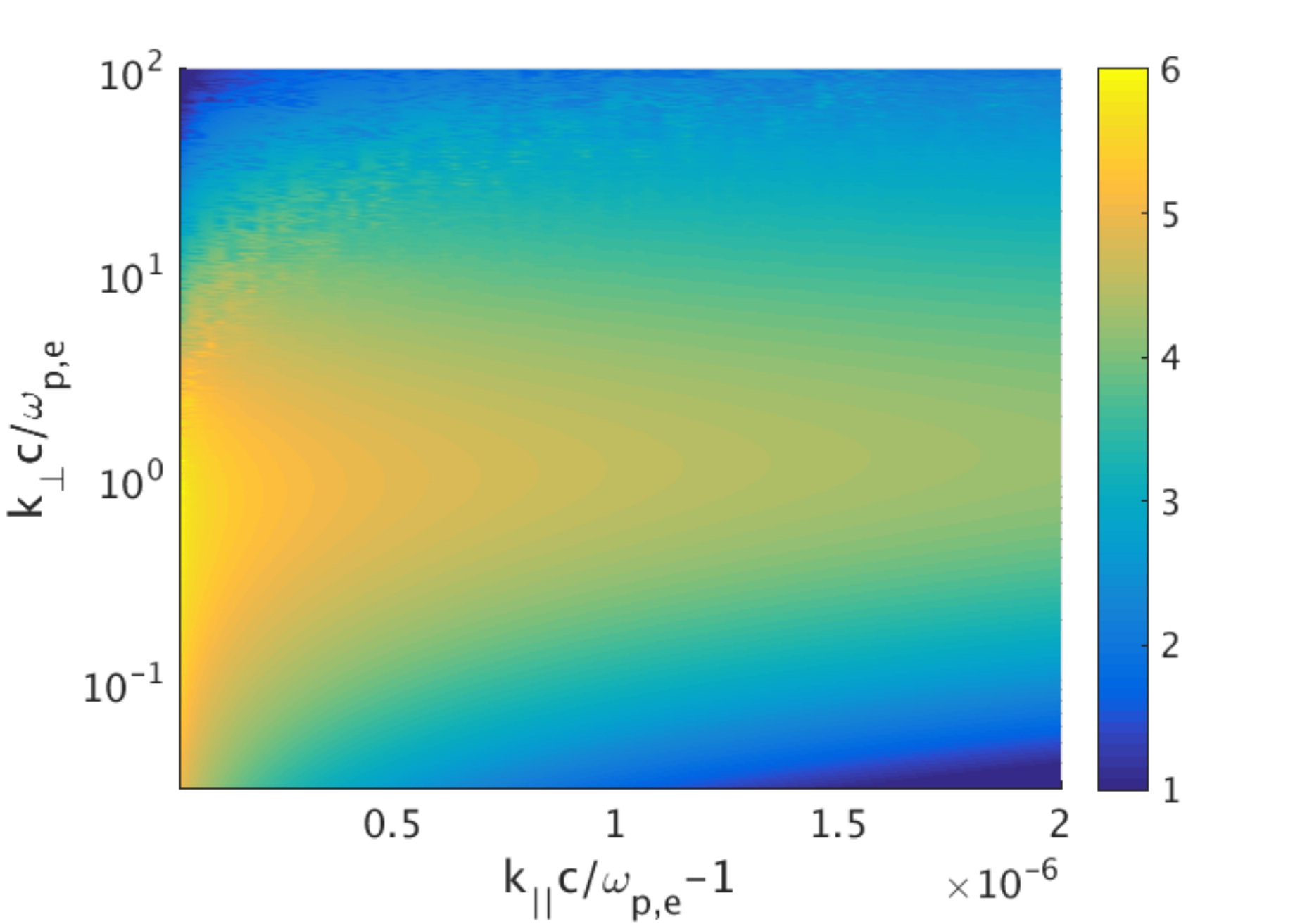}
\caption{ $\log_{10}[\omega_i/(\pi \omega_{p,e}(n_b/n_e))]$ for a blazar-induced pair-beam at the distance 50 Mpc from a blazar (for details see \citet{Vafin18}). $k_\parallel$ and $k_\perp$ denote the wave vector components, respectively, parallel and perpendicular to the beam.}
\label{Log10GrowthRateLargeWaveNumbers}
\end{figure}

The growth rate of the electrostatic instability for an ultra-relativistic beam with the homogeneous number density $n_b$ and the axis-symmetric momentum distribution $f_b(p,\theta)$ in a spherical coordinate system reads

\begin{multline}
\omega_i(k_\perp,k_\parallel)=\pi {n_{b}\over n_e}\omega_{p,e} \l(\omega_{p,e}\over kc\r)^3 \\ \times\int_{\theta_1}^{\theta_2} d\theta 
{-2g\sin\theta + \l( \cos\theta-  (kc/\omega_{p,e}) \cos\theta'\r){\pa g\over\pa\theta} \over
\l[ (\cos\theta-\cos\theta_1) (\cos\theta_2-\cos\theta)  \r]^{1/2}
},
\label{gr1}
\end{multline}
where 
\be
g(\theta)=m_e c \int_0^\infty p f(p,\theta) dp,
\label{gr2}
\ee
\be
\cos\theta_{1,2}={\omega_{p,e}\over kc}\l[ \cos\theta'\pm \sin\theta' \l(  \l(kc\over\omega_{p,e} \r)^2-1\r)^{1/2} \r].
\label{gr3}
\ee
Here, $\theta'$ is the angle between the wave vector ${\bf k}$ and the beam direction, $c$ the speed of light, $n_b$ the beam density, $n_e$ the electron density in the IGM. Using the approximation for the pair-beam distribution at a distance 50 Mpc from a fiducial blazar obtained in our previous work (Eqs. (24)-(25), and (56) from \citet{Vafin18} ), we calculated the electrostatic growth rate Eq. (\ref{gr1}) shown in Fig. \ref{Log10GrowthRateLargeWaveNumbers}. It is well seen that the growth rate decreases towards large $k_\perp$ in contrast to the growth rate used by \citet{Chang14}. Thus, our model provides a more realistic description of the instability growth.

\subsection{Three-dimensional (3D) evolution of the wave spectrum}

Then, the non-linear wave kinetic equation including the LL, NL and collisional damping reads

\be
{dW({\bf k})\over dt}= 2(\omega_i({\bf k}) + \omega_{LL}({\bf k}) +  \omega_{NL}({\bf k})+\omega_c)W({\bf k}),
\label{specteq}
\ee
where $W({\bf k})$ is the spectral energy density of the electric field and  
\be 
\omega_{LL}({\bf k})= - \omega_{p,e} \sqrt{\pi\over 8} \l( \omega_{p,e}\over ku_e \r)^3 
\exp\l[ -{1\over2} \l( \omega_{p,e}\over ku_e \r)^2  \r],
\label{llrate}
\ee

\begin{multline} 
\omega_{NL}= {3(2\pi)^{1/2}\over 64 n_e m_e u_i} \int d^3 k' W({\bf k}') 
{ ({\bf k}{\bf k}')^2\over (k' k)^2} {k'^2-k^2 \over |{\bf k}'-{\bf k}|} \\ \times 
\exp\l[ - a \l( {c\over\omega_{p,e}} {k'^2-k^2 \over |{\bf k}'-{\bf k}|} \r)^2\r],
\label{nlrate}
\end{multline}
are, respectively, the LL and NL damping rates \citep{Breizman72}. Here,  $a=(9/8)[u_e^2/(cu_i)]^2$ where $u_{i,e}=\sqrt{T_\mathrm{IGM}/m_{i,e}}$ denotes the IGM ion and electron thermal speeds and $T_\mathrm{IGM}$ is the IGM temperature. For $T_\mathrm{IGM}=10^4T_4\ \mathrm{K}$: $a\approx3.6\cdot10^{-3}T_4$. {The collisional damping rate reads \citep{Alexandrov84,Miniati13,Huba16}: 
\be 
\omega_c= - 1.45\cdot10^{-6}n_e\lambda T_{e}^{-3/2}\text{ [s$^{-1}$],}
\label{colldamping}
\ee
where $\lambda=23.5-\ln(n_e^{1/2}T_\mathrm{IGM}^{-5/4})-[10^{-5} + (\ln T_\mathrm{IGM} -2)^2/16]^{1/2}$. Here, $n_e$ and $T_\mathrm{IGM}$ are in units cm$^{-3}$ and eV, respectively.} Moreover, we have implicitly assumed that the real part of the frequency $\omega$ is approximately equal to the the plasma frequency $\omega_{p,e}$ in Eqs. (\ref{gr1}) and (\ref{llrate}). It is valid for the wave vectors $kc/\omega_{p,e}\ll c/u_e\approx 0.7\cdot10^3/\sqrt{T_4}$.

The total electric field energy density {is calculated} as 
\be 
W_{tot}= 2\pi\int  W({\bf k}) k_\perp dk_\perp dk_\parallel. 
\label{totalenergy}
\ee
It is to be noted that the NL damping does not affect the total energy density of the electric field. It can be easily seen from the integration of Eq. (\ref{specteq}) over all ${\bf k}$ that the term with the NL damping $\int \omega_{NL}({\bf k})W({\bf k}) dk=0$ actually disappears. Thus, the process, traditionally called NL damping, does not actually damp waves, but only scatters them from larger to smaller wave numbers.   

The spectrum evolution qualitatively consists of several stages. First, the wave energy grows exponentially during the linear instability phase and NL damping can be neglected. Afterward, NL damping starts efficiently scattering the waves towards smaller wave numbers. Then a rapid transition happens when most of the wave energy is transferred out of the instability region in ${\bf k}$-space, and as a consequence, the wave energy approaches a quasi-stationary level. However, the electric field energy continues to increase (but now much slower as compared to the linear phase), since there exists such a region in ${\bf k}$-space where both NL and LL damping are suppressed \citep{Chang14}. In this region, the electrostatic waves are still unstable and can further reduce the beam energy.

Due to the cylindrical symmetry of the problem, the solution $W({\bf k}')$ depends only on $k_\parallel'$ and $k_\perp'$. Then the integral with respect to the angle between ${\bf k}'$ and ${\bf k}$ in Eq. (\ref{nlrate}) can be performed before solving Eq. (\ref{specteq}). This integral is approximately {evaluated} in Appendix \ref{appwnl}. 

As it is seen from Fig. \ref{Log10GrowthRateLargeWaveNumbers}, the electrostatic instability at parallel wave vectors develops in a very narrow band around $k_\parallel c/\omega_{p,e}\approx 1$. Therefore, in the parallel direction, we may limit the problem only to the region $k_\parallel c/\omega_{p,e}\lesssim 1$. 

\begin{figure}
\includegraphics[width = \columnwidth]{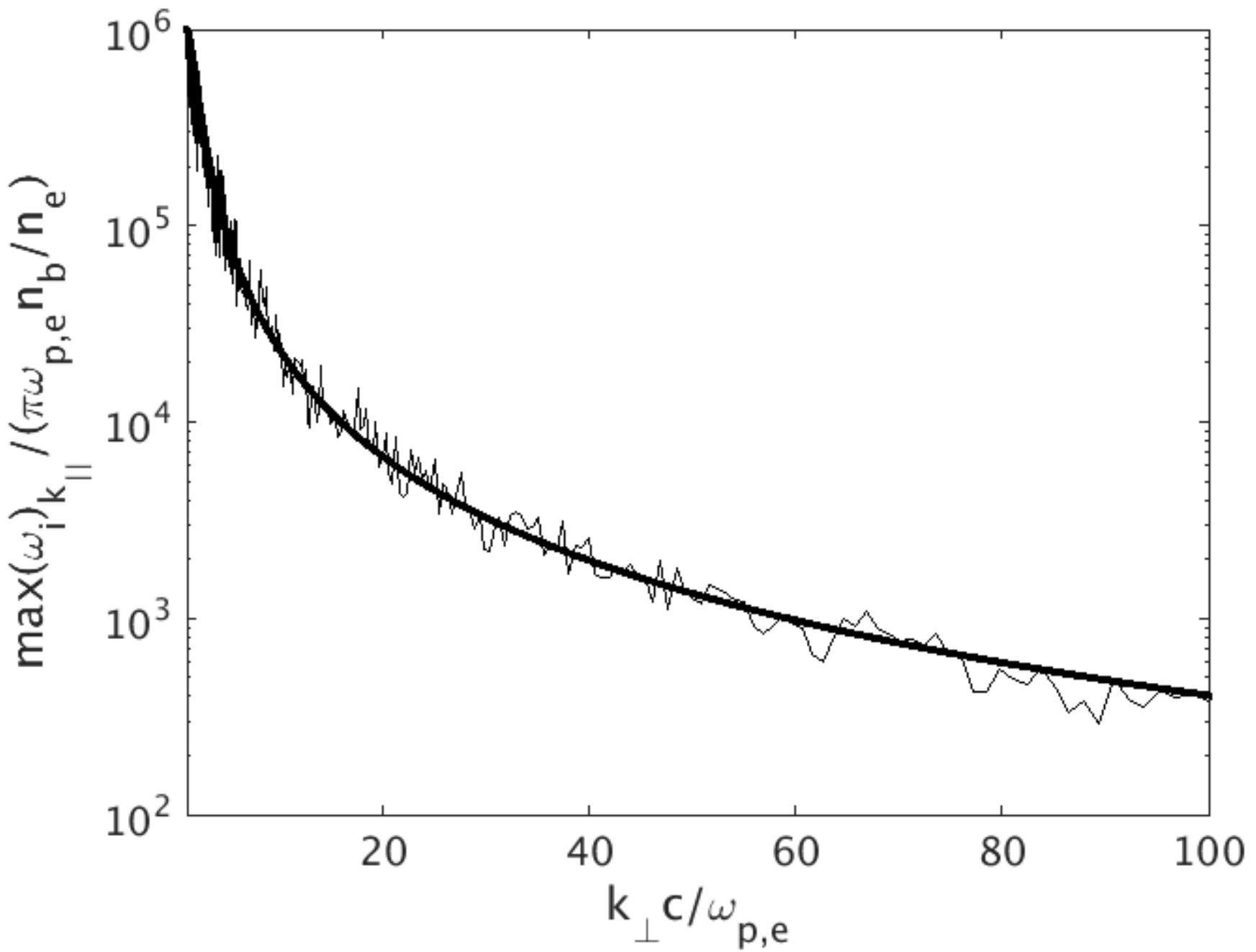}
\caption{ Thin line: the growth rate from Fig. \ref{Log10GrowthRateLargeWaveNumbers} maximized in the parallel direction. Thick line: our approximation Eq. (\ref{gr4}) for the thin line. }
\label{MaxGrowthRatePar}
\end{figure}

\begin{figure}
\hspace{-0.75 cm}
\includegraphics[width = \columnwidth]{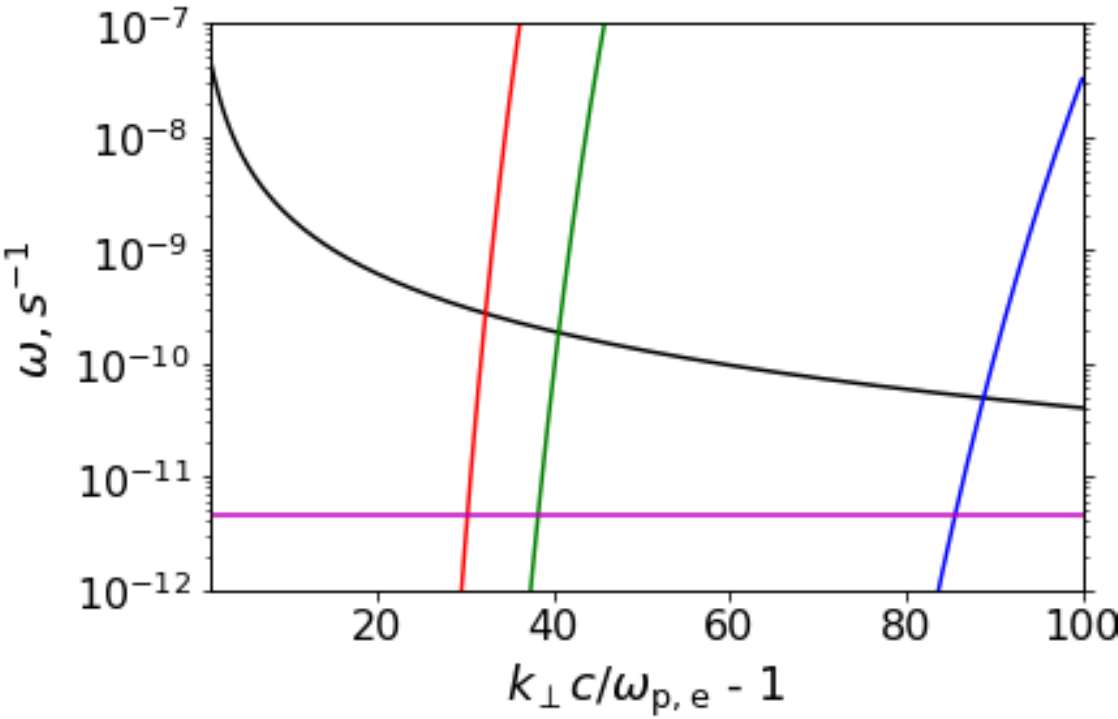}
\caption{Black line: the growth rate Eq. (\ref{gr4}). Magenta line: collisional damping rate at $T_\mathrm{IGM} \approx$ 11000 K, Eq. (\ref{colldamping}). Blue, green and red lines: absolute value of LL damping rate, Eq. (\ref{llrate}) at $T_\mathrm{IGM} \approx$ 11000 K, 58000 K and 93000 K respectively. $n_b=3\cdot10^{-22}$ cm$^{-3}$ and $n_e=10^{-7}$ cm$^{-3}$.}
\label{WiWLL}
\end{figure}

In the perpendicular direction to the beam, the growth rate extends to much {larger} wave numbers, and $k_\perp$ should be considered up to the value where $\omega_{i}\approx\omega_{LL}$. Fig. \ref{MaxGrowthRatePar} demonstrates the growth rate as a function of $k_\perp$, while for each $k_\perp$ we took the maximum growth rate in the parallel direction. The growth rate maximized in the parallel direction can be well approximated by: 

\be
\omega_i(k_\perp)= {\omega_i(k_\perp c/\omega_{p,e}=1) \over 1 + (k_\perp c/\omega_{p,e} )^{1.7} }
\label{gr4}
\ee
which is shown by the thick line in Fig. \ref{MaxGrowthRatePar}. For the pair-beam number density $n_{b}=3\cdot10^{-22}$ cm$^{-3}$ found in our previous work and corresponding to the growth rate presented in Fig. \ref{Log10GrowthRateLargeWaveNumbers} and for the typical IGM parameters $n_e=10^{-7}$ cm$^{-3}$, $T_\mathrm{IGM}=10^4$ K, Fig. \ref{WiWLL} compares the growth rate Eq. (\ref{gr4}) with the LL {and collisional damping rate, Eq. (\ref{llrate}) and Eq. \ref{colldamping}}. Since $k_\parallel c/\omega_{p,e}\lesssim 1$ and the LL rate is significant at $kc/\omega_{p,e}\gg1$, we may set $k\approx k_\perp$ in Eq. (\ref{llrate}). Fig. \ref{WiWLL} shows that the LL damping suppresses the electrostatic instability at $k_\perp c/\omega_{p,e}\gtrsim 90$, so we may take into account only $k_\perp c/\omega_{p,e}\lesssim 100$. For higher temperatures, say $T_\mathrm{IGM} \approx$ $6\cdot10^4$ K and $9\cdot10^4$ K, the LL damping suppresses the {electrostatic} instability at $k_\perp c/\omega_{p,e}\gtrsim 45$ and $k_\perp c/\omega_{p,e}\gtrsim 35$, respectively. The range of $k_\perp c/\omega_{p,e}$, for our simulations, is chosen accordingly.

As the initial condition for Eq. (\ref{specteq}), the constant value given by the discrete particle fluctuations can be used \citep{Klimontovich82,RS12}: 
\be
W({\bf k},t=0)= W_0 \approx k_BT_\mathrm{IGM} \approx 1.6\cdot10^{-12} T_{4},\text{ erg},
\label{thermfluct}
\ee 
where $T_\mathrm{IGM}=10^4 T_{4}$ K. Eq. (\ref{thermfluct}) provides the noise level of electric field turbulence during the time evolution. As there is no term responsible for the noise generation in Eq. (\ref{specteq}), the solution $W({\bf k})$ must be forced to stay above $W_0$ by the condition $W({\bf k})\geq W_0$ in couple with Eq. (\ref{specteq}). However, this provides an artificial energy source, but it contributes a much smaller energy fraction compared to the effect of electrostatic instability.         

\subsection{Reduction to two-dimensional (2D) model}

To investigate Eq. (\ref{specteq}), we utilize a 2D model ($W=W(k_\perp)$) for analyzing the problem in the plane $k_\parallel c/\omega_{p,e}\approx 1$ of the ${\bf k}$-space. To preserve the fact that the waves are scattered out to the non-resonant wave vectors, i.e., out of the instability region, we modify the growth rate to 

\begin{multline}
\omega_i(k_\perp)= {\omega_i(k_\perp c/\omega_{p,e}=1)\theta(k_\perp c/\omega_{p,e}-1) \over 1 + (k_\perp c/\omega_{p,e} )^{1.7} }\approx \\
5.6\cdot10^{-6} \frac{n_{b20}}{\sqrt{n_{e7}}} \frac{\theta(k_\perp c/\omega_{p,e}-1)}{1+(k_\perp c/\omega_{p,e})^{1.7}},\text{ s$^{-1}$},
\label{grmod}
\end{multline}
where $\theta$ is the Heaviside step function. The growth rate is truncated at $k_\perp c/\omega_{p,e} = 1$, for waves at small $k_\perp$ to be stable. At the same time, Eqs. (\ref{llrate})-(\ref{nlrate}) and Eq. (\ref{totalenergy}) can be reduced to
\be 
\omega_{LL}= - \omega_{p,e} \sqrt{\pi\over 8} \l( \omega_{p,e}\over k_\perp u_e \r)^3 
\exp\l[ -{1\over2} \l( \omega_{p,e}\over k_\perp u_e \r)^2  \r],
\label{llrate1d}
\ee

\begin{multline} 
\omega_{NL}= {3(2\pi)^{1/2}\over 64 n_e m_e u_i} \int d^2 k_\perp' W( k_\perp',t) 
{ ({\bf k_\perp}{\bf k_\perp}')^2\over (k_\perp' k_\perp)^2} {k_\perp'^2-k_\perp^2 \over |{\bf k}_\perp'-{\bf k}_\perp|} \\ \times 
\exp\l[ - a \l( {c\over\omega_{p,e}} {k_\perp'^2-k_\perp^2 \over |{\bf k}_\perp'-{\bf k}_\perp|} \r)^2\r],
\label{nlrate1d}
\end{multline}

\be 
W_{tot}(t)= 2\pi\int  W( k_\perp,t) k_\perp dk_\perp.
\label{totalenergy1d}
\ee
%where $\Delta k_\parallel\approx 10^{-6}\omega_{p,e}/c$ is the characteristic width of the instability region parallel direction. 
The angular integral in Eq. (\ref{nlrate1d}) can be evaluated similar to Appendix \ref{appwnl} with the result presented in Appendix \ref{appwnl1d}. 

The initial condition for Eq. (\ref{specteq}) in 2D can be taken as
\be
W( k_\perp,t=0)= W_{0, \perp} \approx k_BT_\mathrm{IGM}\Delta k_\parallel ,
\label{thermfluct1d}
\ee 
where $\Delta k_\parallel\approx 10^{-6}\omega_{p,e}/c$ is approximately the width of the instability region in the parallel direction (see Fig. \ref{Log10GrowthRateLargeWaveNumbers}). Eq. (\ref{thermfluct1d}) can be a rather rough estimate, but the exact estimation of the noise energy density is unimportant for the later evolution as pointed out by \citet{Chang14}. {Appendix \ref{applicability2D} demonstrates analytically that the 2D model reflects the main features of the original 3D one and can be used to study the spectral equation (\ref{specteq}). The results of a 3D test simulation are also discussed in Appendix \ref{comparison3D2D} in comparison to those of our 2D model.

\section{Numerical results {without collisions}}\label{results}

{We first analyze the system of Eqs. (\ref{specteq}) and (\ref{grmod})-(\ref{nlrate1d}) numerically using the finite-difference method and neglecting particle collisions}. The pair beam density and average gamma-factor are, respectively, $n_b=3\cdot10^{-22}$ cm$^{-3}$ and $<\gamma_b>=4\cdot10^6$ that correspond to the pairs produced within the IC mean free path at distance $50$ Mpc from a fiducial source of high-energy gamma-rays \citep{Vafin18}. For this beam setup, we then consider the effect of the IGM temperature on the energy loss time of the beam $\tau(\alpha)$ defined as
\be
\int_0^{\tau(\alpha)} P(t) dt=\alpha W_b(t=0)
\label{tau}
\ee
where 
\be 
P(t)= 8\pi\int \omega_i W(k_\perp,t) k_\perp dk_\perp
\label{powerloss}
\ee
is the power loss of the beam, $W_b(t=0)=<\gamma_b>n_bm_ec^2$ the initial beam energy density, and $\alpha$ the fraction of initial beam energy lost at time $\tau(\alpha)$. Here, we account for the equipartition of beam energy into kinetic energy and electrostatic energy of the plasma waves. In the case of a cold beam, the fraction $\alpha$ can be as high as 0.5 \citep{RS02} which we will also use for our estimation of the relaxation time of the beam $\tau_{rel}=\tau(0.5)$. 

In principle, the relaxation time can be estimated from the simulation. However, if $\tau_{rel}$ is large, it can take a prohibitively long time to achieve it numerically, even in 2D. A more efficient way of calculating $\tau_{rel}$ is to extrapolate simulation results in the following manner. It will be demonstrated below that during the quasi-saturation stage the power loss function $P(t)$ changes almost periodically. Since the system turns to a quasi-saturation at a time $t_{NL}$ much shorter than $\tau_{rel}$ and the beam energy loss at $t=t_{NL}$ is only about 1\%, we  can evaluate the integral Eq. (\ref{tau}) approximately as
\be 
\int_0^{\tau(\alpha)} P(t) dt\approx  {\tau\over T}\int_{t_{NL}}^{t_{NL}+T} P(t) dt,
\label{apptau}
\ee
where $T$ is the oscillation period of $P(t)$ at $t>t_{NL}$. Then
\be 
\tau_{rel}= {T W_b(t=0)\over2} \l(  \int_{t_{NL}}^{t_{NL}+T} P(t) dt \r)^{-1}.
\label{trel}
\ee

The simulation parameters are summarized in Table \ref{Table1}. Case 1 treats the beam evolution in the IGM with the average parameters, while cases 2 and 3 consider higher temperatures of the IGM. The energy loss time becomes shorter for a higher IGM temperature, because of smaller NL damping. The latter has to do with the exponential cut-off in integrand of $\omega_{NL}$ where the factor $a\propto T_e$. Thus, the NL damping is weaker for higher plasma temperatures. {Cases 2 and 3} with higher IGM temperatures refer to the IGM heating scenario by blazar-induced pair beams proposed by \citet{Puchwein12}. Therefore, it is crucial to understand qualitatively how the IGM temperature effects the relaxation time of the pair beam.

The highest value of $k_\perp$ in each simulation is determined by equality of the LL damping rate and the linear growth rate (see Fig. \ref{MaxGrowthRatePar}). The lowest value of $k_\perp$ was chosen small enough ($10^{-8}\omega_{p,e}/c$) to prevent peaking of the solution at the single left grid point within the simulated time interval. The number of grid points was adjusted automatically at each time step to maintain the accuracy of the solution within 10 \%. 

\begin{table}
\caption{Simulation parameters and relaxation time} % title of Table
\centering % used for centering table
\begin{tabular}{|c|c|c|c|} % centered columns (4 columns)
\hline %inserts double horizontal lines
Case & $n_e$, cm$^{-3}$ & $T_{e}$, eV & $\tau_{rel}$, s \\ [0.5ex] 
\hline %horizontal line
1 & $10^{-7} $ & 1 & $7\cdot10^{15}$ \\
\hline
2 & $10^{-7}$ & 5 & $7\cdot10^{13}$ \\
\hline
3 & $10^{-7}$ & 8 & $1.6\cdot10^{13}$ \\
\hline
\end{tabular}
\label{Table1} % is used to refer this table in the text
\end{table}

Fig. \ref{Wkcase1} illustrates the spectrum evolution for case 1. During early times $t\lesssim 10^8$ s, there is one pronounced maximum in the spectrum (black curve in Fig. \ref{Wkcase1}) resulting from the modes with the highest growth rate. Afterward, NL damping becomes important and efficiently scatters waves to lower $k_\perp$. As a result, the peak at $k_\perp c/\omega_{p,e}\sim 1$ disappears, and the spectral density in the region of the highest growth rate $W(k_\perp c/\omega_{p,e}\sim1)$ becomes equal to the noise level $W_{0,\perp}$.  Any subsequent growth at $k_\perp c/\omega_{p,e}\sim 1$ is efficiently suppressed by NL damping, and W($k_\perp c/\omega_{p,e}\sim 1$) remains at the noise level during the remainder of the simulation time. In this case, the NL damping is determined by the non-resonant waves at $k_\perp c/\omega_{p,e}<1$.  

Therefore, as mentioned above, the long-term beam evolution is determined by the spectral properties at large $k_\perp$, where both the LL and NL damping rates can be smaller than the growth rate. Fig. \ref{Wkcas1LargeK} demonstrates the spectrum at large $k_{\perp}$ for much longer simulation times compared to Fig. \ref{Wkcase1}. The spectrum at large $k_\perp$ evolves somewhat periodically: the waves continuously grow and get scattered to smaller wave numbers. The right maximum in Fig. \ref{Wkcas1LargeK} corresponds to the region where the waves are amplified and the left one arises due to the NL damping of the waves in the right peak. NL damping transfers this energy to the low-$k$ region at later times. However, the saturation energy density at low-$k_{\perp}$ remains effectively constant since the energy density at large $k_{\perp}$ is much smaller compared to the energy of non-resonant waves at $k_{\perp}c/\omega_{p,e} < 1$. The spectrum at $k_\perp c/\omega_{p,e}<1$ stays almost constant, and the spectrum exhibits a flat peak at $k_\perp\simeq0$.

\begin{figure}
\includegraphics[width = \columnwidth]{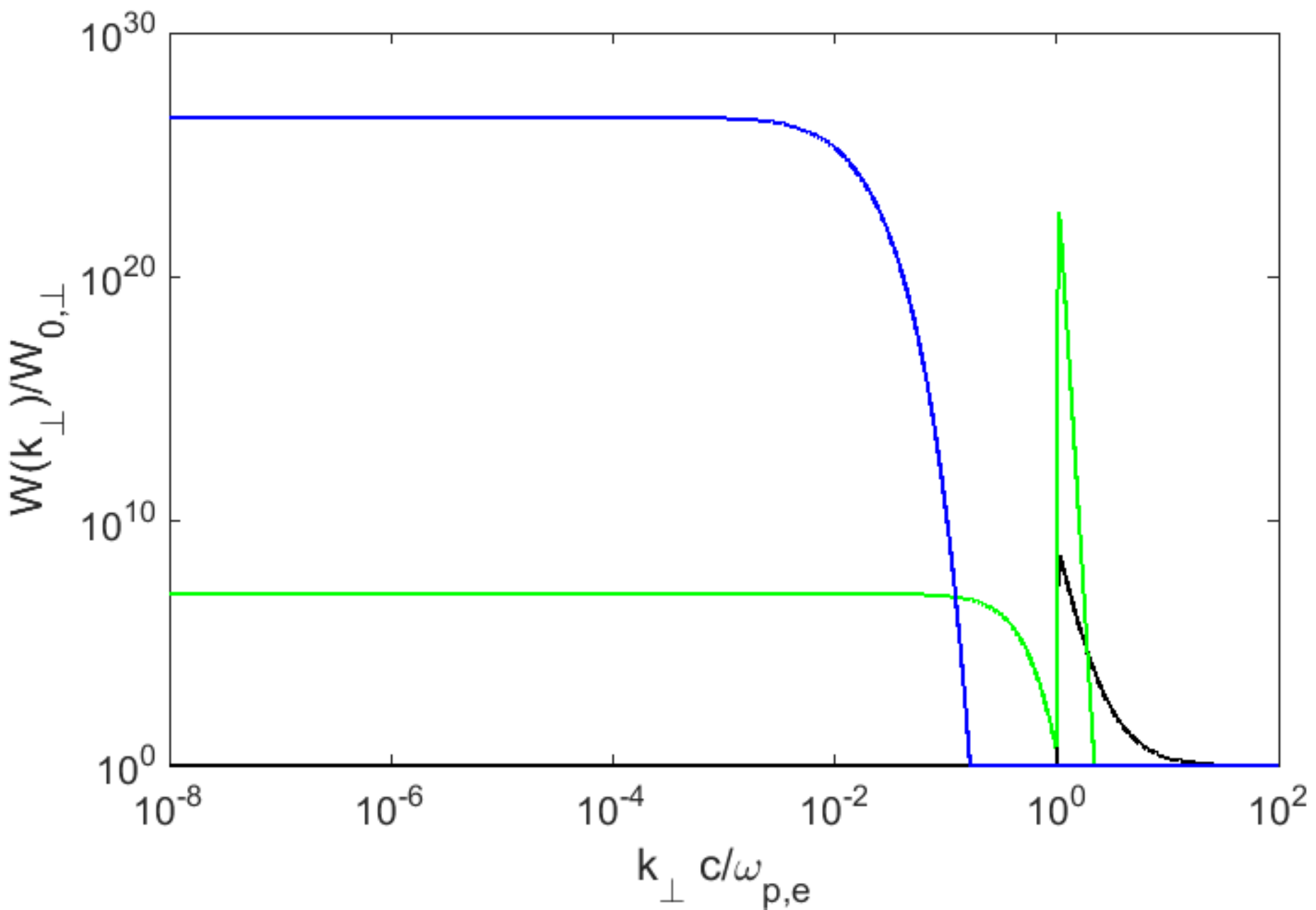}
\caption{The spectrum evolution for case 1. Black: $t=1.2\cdot10^8$ s. Green: $t=3.2\cdot10^8$ s. Blue: $t=3.7\cdot10^8$ s.}
\label{Wkcase1}
\end{figure}

\begin{figure}
\includegraphics[width = \columnwidth]{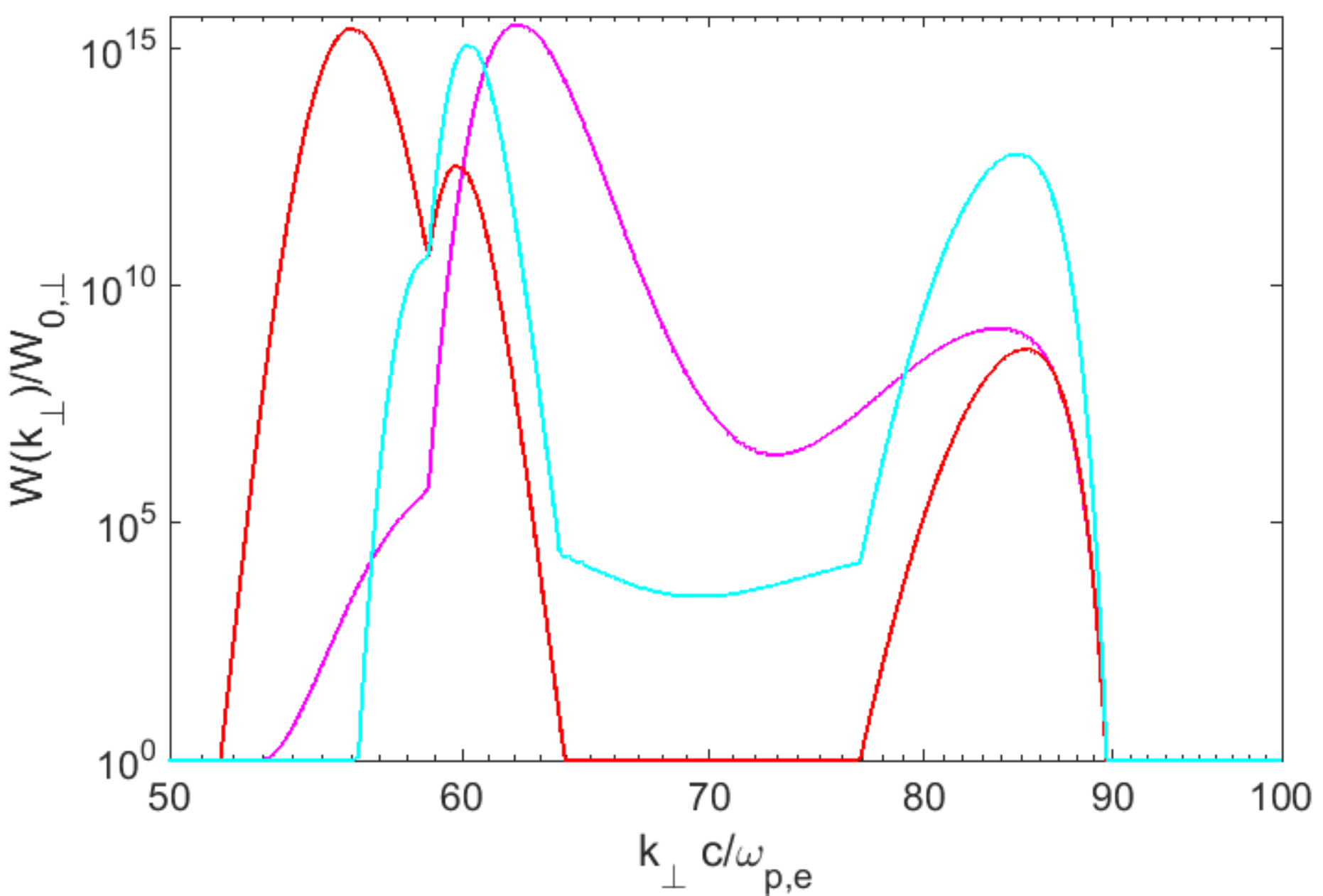}
\caption{The spectrum evolution for case 1 at large $k_{\perp}$. Magenta: $t=1.3\cdot10^{11}$ s. Red: $t=1.4\cdot10^{11}$ s. Cyan: $t=2\cdot10^{11}$ s.}
\label{Wkcas1LargeK}
\end{figure}

The evolution of the total energy density of the electric field and the power losses of the beam for case 1 are shown in Figs. \ref{Wcase1} and \ref{Pcase1}, respectively. It is well seen from Fig. \ref{Wcase1} that the beam loses only a small fraction of its energy ($\sim10^{-2}W_b$) during the linear instability phase $t\lesssim3\cdot10^8$ s. At the same time, $P(t)$ achieves its highest value at $t\approx3.3\cdot10^8$ s. However, almost immediately afterward the power losses undergo a rapid decline due to the fact that most of the waves are scattered to the region $k_\perp c/\omega_{p,e}<1$ where the instability growth rate is zero, $\omega_i=0$, in our model. The wave growth at large $k_\perp$ results in the increase of $P(t)$ again at $t>10^{10}$ s. Eventually, $P(t)$ turns to a periodic solution.

Fig. \ref{Pcases13} compares the power losses of the beam for cases 1-3. Here, it is well seen that the function $P(t)$ changes periodically during the quasi-saturation phase and, therefore, our extrapolation of the time integral given by Eq. (\ref{apptau}) is justified. However, since the oscillation amplitude of $P(t)$ changes with time, we considered a time interval $T$ where $P(t)$ achieves the maximum multiple times to estimate Eq. (\ref{apptau}). We also note that the power losses are strongly affected by the plasma temperature. The power losses grow with the increase in plasma temperature because the NL damping rate becomes suppressed due to a stronger exponential cutoff in the NL damping rate as evident from Fig. \ref{Pcases13}. The relaxation time Eq. (\ref{trel}) for cases 1-3 is summarized in Table \ref{Table1}. The relaxation time should be compared to the IC scattering time $\tau_{IC}\approx 10^{20}/\gamma_b=2.5\cdot10^{13}$ s for $\gamma_b=4\cdot10^6$. We conclude that only for a hot enough IGM the relaxation time can become slightly smaller than the IC time.

Thus, if the NL damping is the only non-linear stabilization process, the electrostatic instability has a negligible effect on the beam energy dissipation within the IC time scale, $\tau_{IC}$, and other effects should be invoked to explain the lack of the cascade GeV signal \citep{Neronov10}. The reason for such a strong suppression of the instability is that the NL damping rate in the region with the highest growth rate ($k_\perp c/\omega_{p,e}\simeq 1$) is proportional to the energy density of the long-wave non-resonant oscillations at $k_\perp c/\omega_{p,e}\ll 1$. Obviously, an effective dissipation of energy in the long-wave part of the spectrum will result in a smaller NL damping rate and, consequently, in a shorter energy loss time of the pair beam modifying the conclusion of our simulations above. Possible dissipation mechanisms are the modulation instability and collisional damping. While the former requires a thorough investigation going well beyond the scope of the current work, the collisional rate can be easily estimated for electrons with $n_e=10^{-7}$ cm$^{-3}$ and $T_e=1$ eV to be $\nu_c\approx 10^{-11}$ s$^{-1}$. Since the IC time for the beam considered above is $\tau_{IC}\approx 2.5\cdot10^{13}$ s, the collisional dissipation can significantly affect the value of $\tau_{loss}$ in case 1. It is to be noted that for case 3, where $\tau_{loss}<\tau_{IC}$ s, $\nu_c\approx 2.2\cdot10^{-13}$ s and collisions do not play a considerable role at $t\lesssim\tau_{IC}$.

\begin{figure}
\includegraphics[width = \columnwidth]{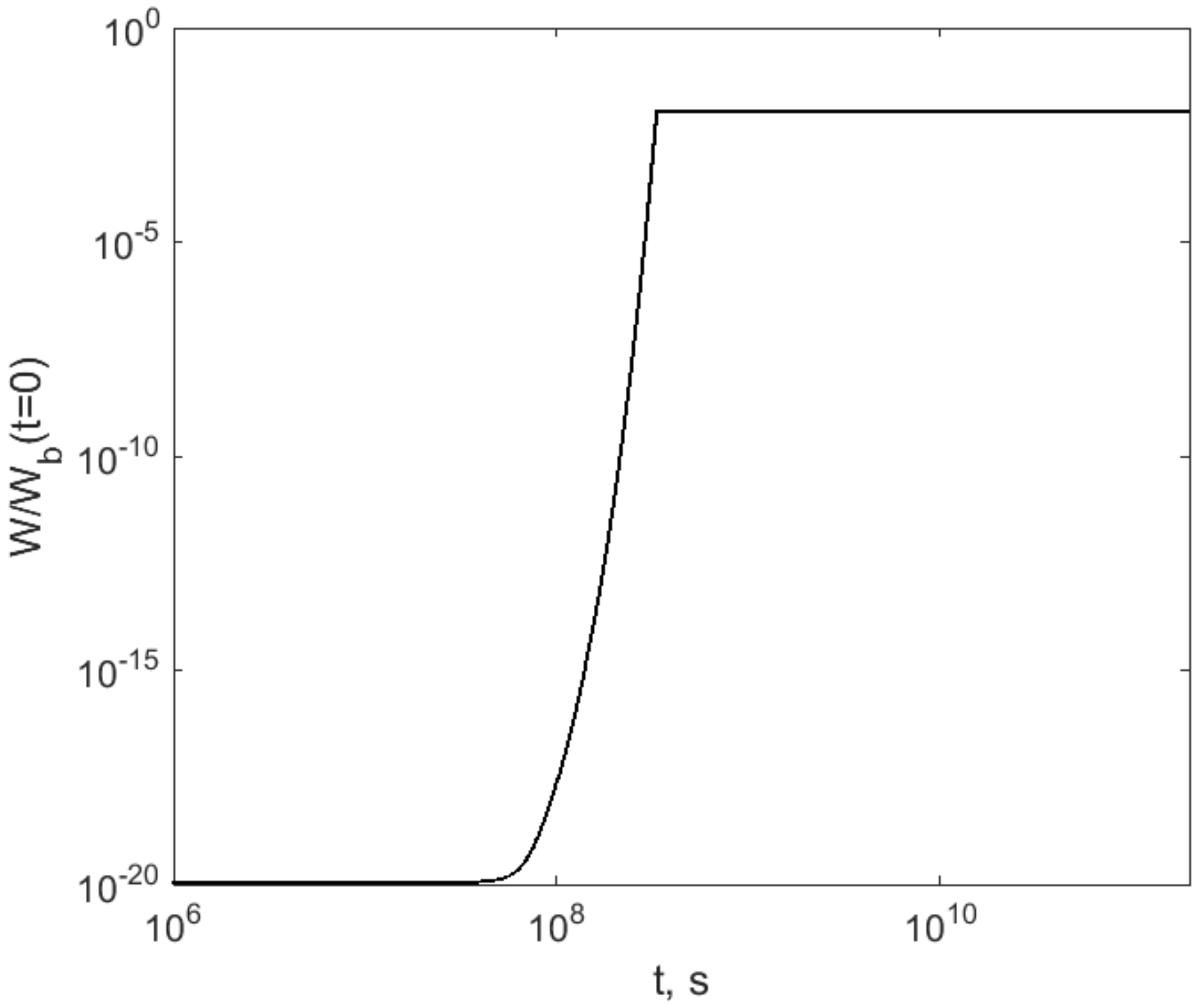}
\caption{Total electric field energy density for case 1. $W_b(t=0)=<\gamma_b>n_b m_ec^2$.}
\label{Wcase1}
\end{figure}

\begin{figure}
\includegraphics[width = \columnwidth]{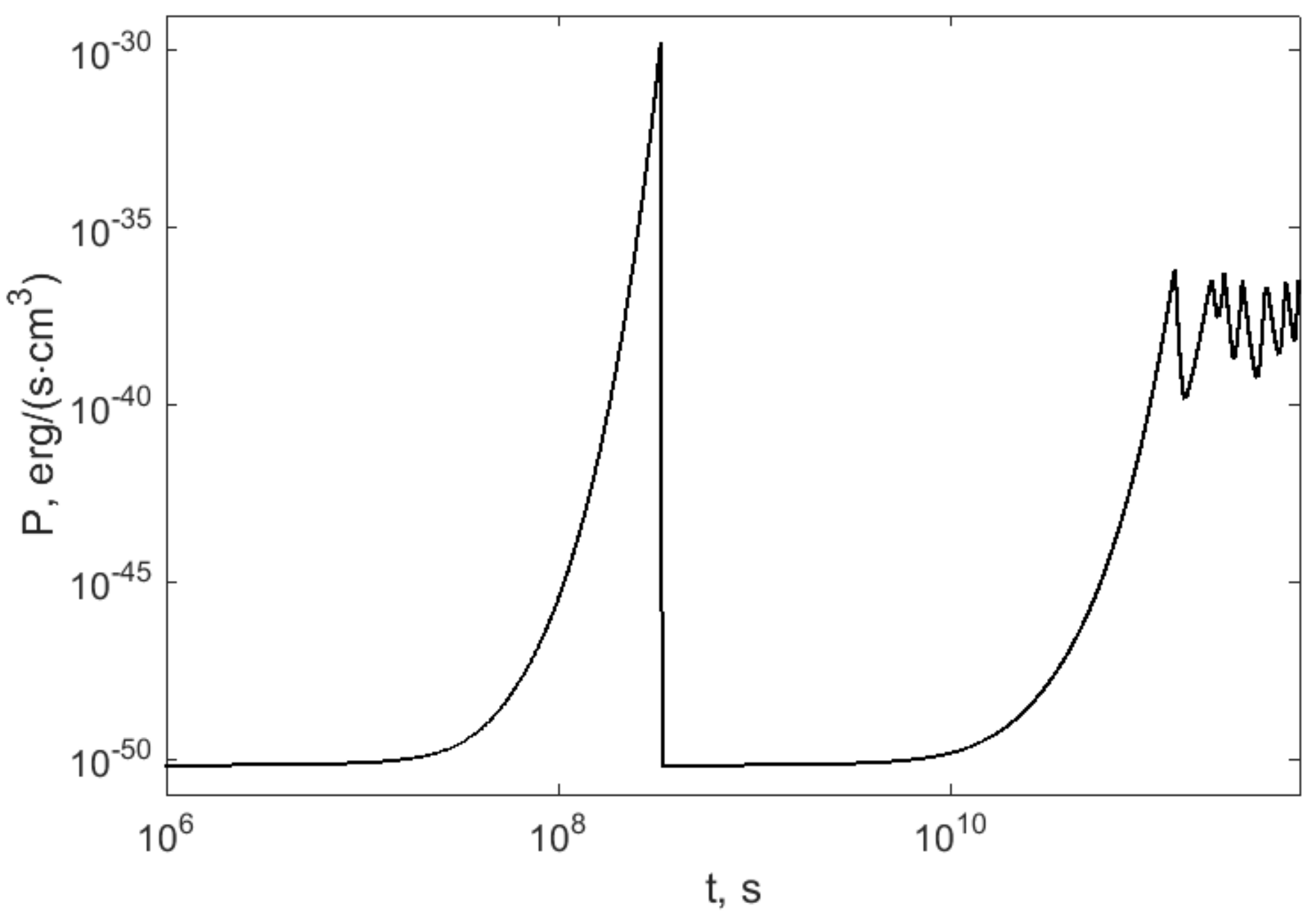}
\caption{Power losses of the pair beam, Eq. (\ref{powerloss}), for case 1. }
\label{Pcase1}
\end{figure}

\begin{figure}
\includegraphics[width = \columnwidth]{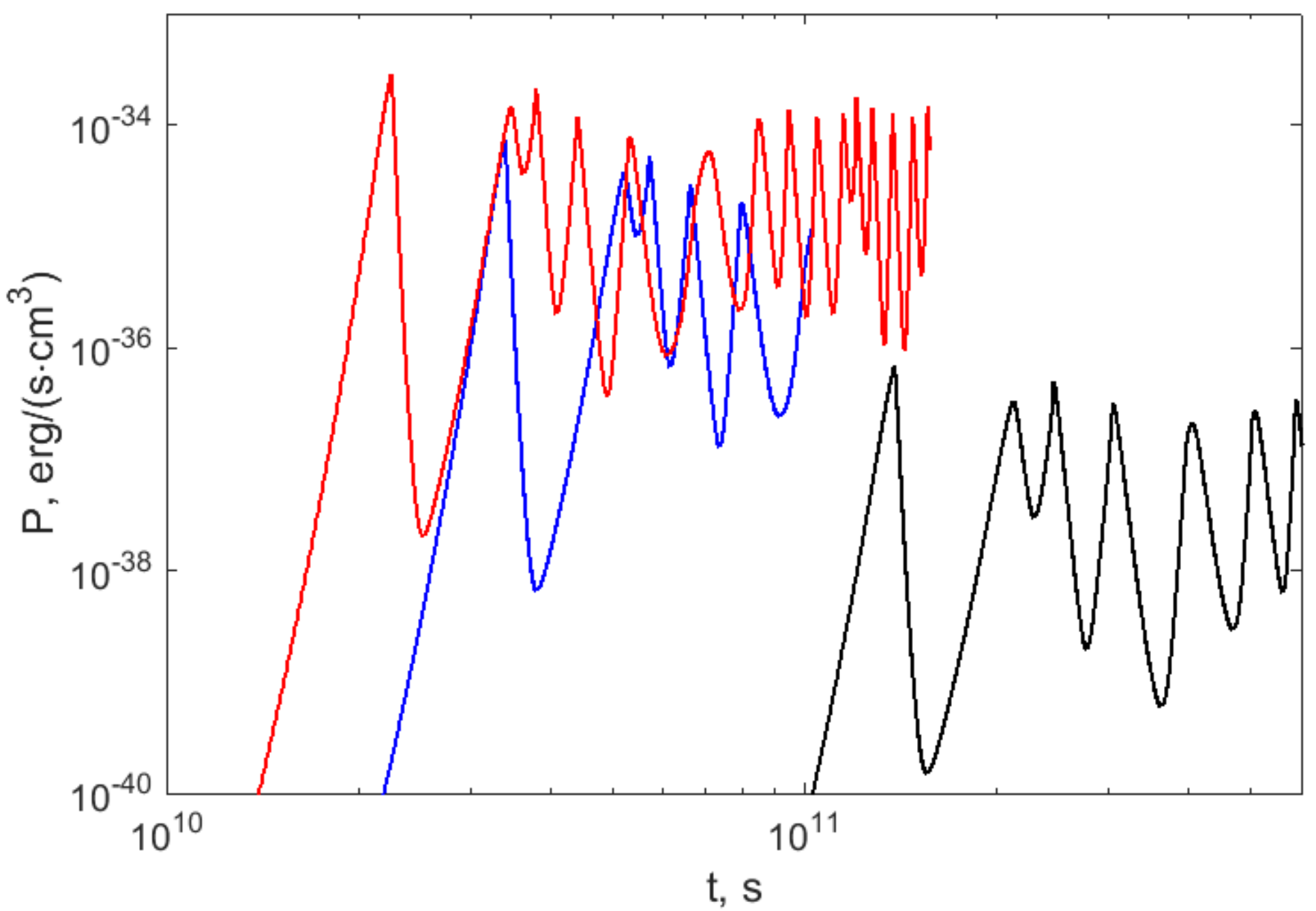}
\caption{Power losses of the pair beam, Eq. (\ref{powerloss}), for case 1(black), 2(blue), and 3(red).}
\label{Pcases13}
\end{figure}

\section{Effect of particle collisions}\label{collisions}

Presence of particle collisions results in an additional damping rate in Eq. (\ref{specteq}).  In view of the fact that the effect of particle collisions becomes important for low IGM temperatures, we carried out three simulations with parameters given by Table \ref{Table2}. 

%We will see below that presence of collisions affects, first of all, spectrum evolution at $k_\perp c/\omega_{p,e}\sim1$. 
%At the same time, case 1 ($T_e=1$ eV) above clearly demonstrates that in the absence of collisions the relaxation time is determined by the waves at large $k_\perp c/\omega_{p,e}\gg1$. However, this relaxation time is much larger compared to $\tau_{IC}$. For smaller plasma temperatures ($T_e<1$ eV), the last conclusion holds as well. We will demonstrate below that case 4 (which is basically case 1 but including collisions) provides a relaxation time much smaller than case 1. For cases 5-6, this discrepancy should be even larger as the collision rate is greater. Thus, the region $k_\perp c/\omega_{p,e}\sim1$ for cases 4-6 is decisive for the beam relaxation time. As a result, we can consider smaller wave number range $k_\perp c/\omega_{p,e}\leq10$ to study the effect of collisions (cases 4-6).       

\begin{table}
\caption{Simulations including particle collisions} % title of Table
\centering % used for centering table
\begin{tabular}{|c|c|c|c|} % centered columns (4 columns)
\hline %inserts double horizontal lines
Case & $n_e$, cm$^{-3}$ & $T_{e}$, eV & $\tau_{rel}$, s \\ [0.5ex] 
\hline %horizontal line
4 & $10^{-7} $ & 1 & $2.7\cdot10^{14}$ \\
\hline 
5 & $10^{-7}$ & 0.8 & $2.2\cdot10^{14}$ \\
\hline
6 & $10^{-7}$ & 0.3 & $8.3\cdot10^{13}$ \\
\hline
\end{tabular}
\label{Table2} % is used to refer this table in the text
\end{table}

Let us consider case 4 which treats the same IGM parameters as case 1 but now including the effect of particle collisions. In this case, collisions are negligible up to the time $\approx 10^{10}$ s, but afterward, they considerably dissipate the electric field energy by almost two orders of magnitude which is illustrated in Fig. \ref{W_case1_4}. If we assume for a rough estimation that collisions dissipate the energy exponentially $\propto\exp(-2|\omega_c| t)$, then the energy will be reduced by two orders of magnitude after the time $\Delta t=\ln(10)/|\omega_c|\approx 2.3/|\omega_c|\approx5.1\cdot10^{11}$ s which perfectly agrees with our numerical result.} Collisions reduce the energy density of the non-resonant waves at small $k_\perp c/\omega_{p,e}<1$, as the collision frequency can be neglected compared to the linear growth rate at $k_\perp c/\omega_{p,e}>1$ (see Fig. \ref{WiWLL}). This energy dissipation modifies the NL damping rate which is now also reduced by almost two orders of magnitude (see Fig. \ref{Comp_wNL_case1_4}). As a result, the NL damping rate can become smaller than the maximum instability growth rate near $k_\perp c/\omega_{p,e}=1$ as it is illustrated in the zoomed view in Fig. \ref{Comp_wNL_case1_4}. It naturally leads to an electric field growth at $k_\perp c/\omega_{p,e}=1$ at later times in addition to the electrostatic waves at large $k_\perp$. The corresponding spectrum of these waves is shown in Fig. \ref{Wkcase1_coll} for several instants of time (we do not show the region $k_\perp c/\omega_{p,e}\gg1$ because it remains similar to case 1, Fig. \ref{Wkcas1LargeK}). It is to be emphasized that a similar peak at later times did not appear in case 1, as the NL damping rate at $k_{\perp}c/\omega_{p,e}=1$ for case 1 stays above the growth rate during the non-linear stage. The spectral peak in Fig. \ref{Wkcase1_coll} periodically grows and drops with time which drastically modifies the power losses of the beam. Fig. \ref{Pcase4} shown the time evolution of the power losses for case 4 at later times when the effect of collisions becomes noticeable. In Fig. \ref{Pcase4}, we can recognize two periodic structures: sharp large amplitude and broad small amplitude peaks. The latter is similar to case 1 and is caused by the electric field at $k_\perp c/\omega_{p,e}\gg1$, while the former results is an effect of collisions and is related to the spectral peak at $k_\perp c/\omega_{p,e}=1$ (Fig. \ref{Wkcase1_coll}). Due to collisions, the power losses achieve a much higher value compared to case 1. This reduces the relaxation time by factor 20 and it becomes $2.7\cdot10^{14}$ s. Thus, the beam losses its energy much faster (Fig. \ref{Wbt}). Table \ref{Table2} compares the relaxation time for smaller IGM temperatures. From what follows that the relaxation time is well above the IC time for small IGM temperatures even if collisions are present. 

%Fig. \ref{Wkcase1_coll} demonstrates the spectrum evolution at $k_\perp c/\omega_{p,e}\sim1$ for case 4. The spectrum exhibits a sharp peak (absent in case 1) which periodically grows and drops with time which results in oscillating power losses of the beam qualitatively similar to case 1. However, compared to case 1, the power losses in case 4 achieve much higher value (see Fig. \ref{Pcase4}). Eq. (\ref{trel}) yields the relaxation time $2.7\cdot10^{14}$ which is about 20 times smaller compared to the case without collisions (case 1). Since the relaxation time for case 4 is much smaller compared to case 1, it results in a more pronounced decrease of the beam energy density (Fig. \ref{Wbt}). Table \ref{Table2} also the relaxation times for smaller IGM temperatures. The relaxation time is, however, well above the IC time.  

\begin{figure}
\includegraphics[width = \columnwidth]{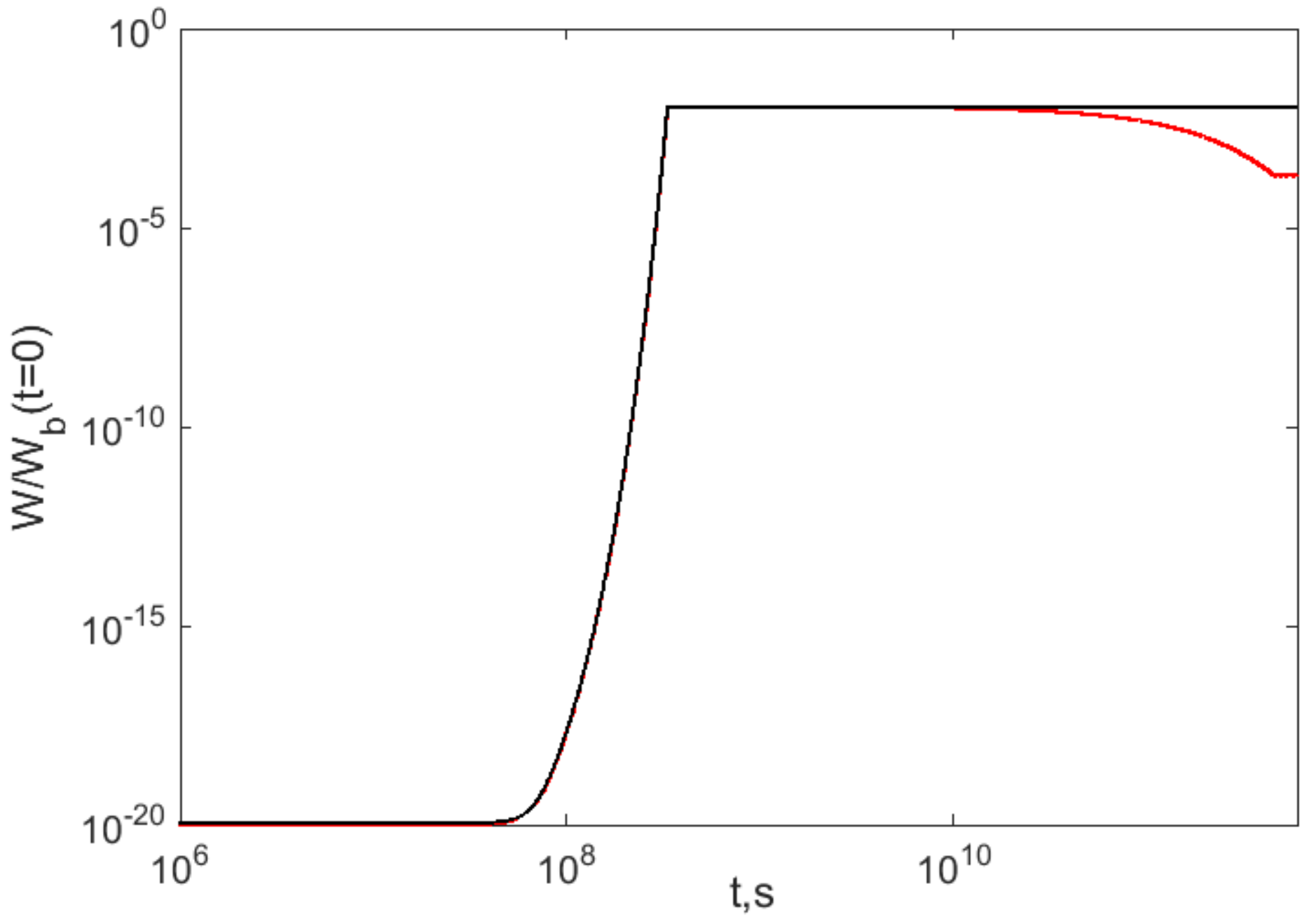}
\caption{Comparison of the total electric field energy density for case 1(black curve) and case 4(red curve). $W_b(t=0)=<\gamma_b>n_b m_ec^2$.}
\label{W_case1_4}
\end{figure}

\begin{figure}
\includegraphics[width = \columnwidth]{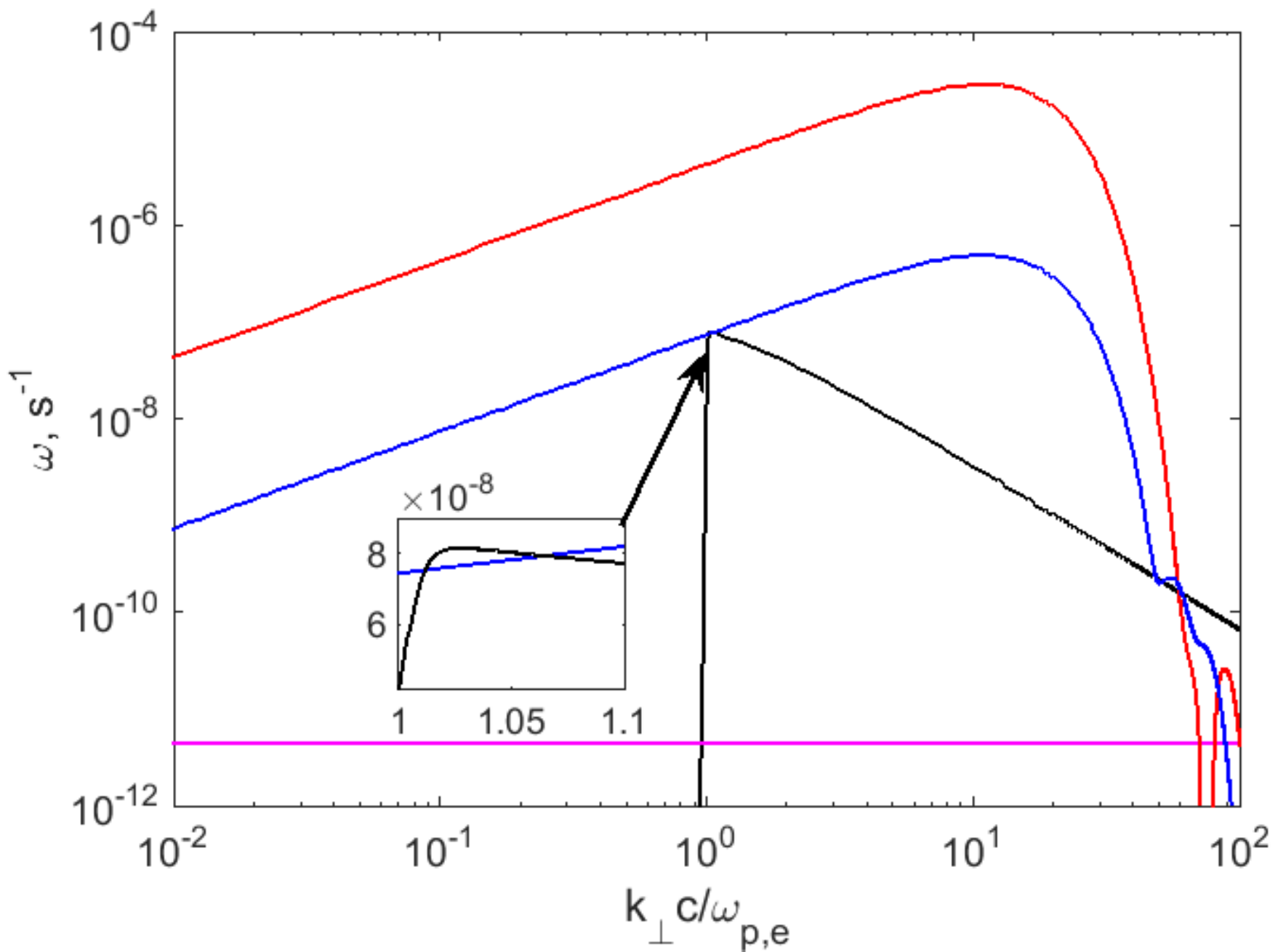}
\caption{Comparison of the NL damping rate (taken with opposite sign) for case 1 (red curve) and case 4(blue curve) at time $t=5.607\cdot10^{11}$ s. The black and magenta lines illustrate, accordingly, the linear electrostatic growth rate Eq. (\ref{grmod}) and the collisional damping rate (\ref{colldamping}) (taken with the opposite sign).}
\label{Comp_wNL_case1_4}
\end{figure}

\begin{figure}
\includegraphics[width = \columnwidth]{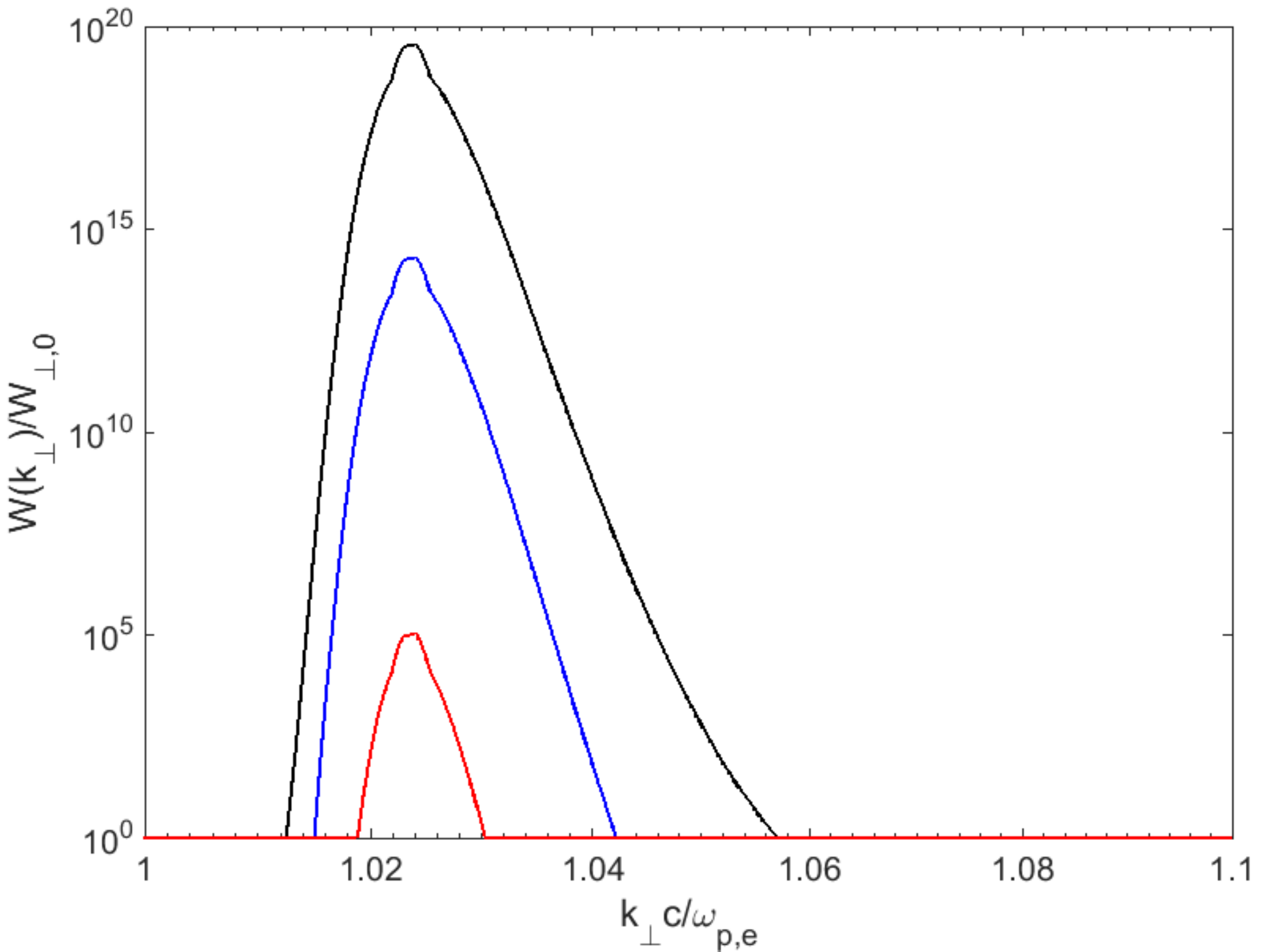}
\caption{The spectrum evolution for case 4. Black: $t=5.616\cdot10^{11}$ s. Blue: $t=5.628\cdot10^{11}$ s. Red: $t=5.656\cdot10^{11}$ s.}
\label{Wkcase1_coll}
\end{figure}

\begin{figure}
\includegraphics[width = \columnwidth]{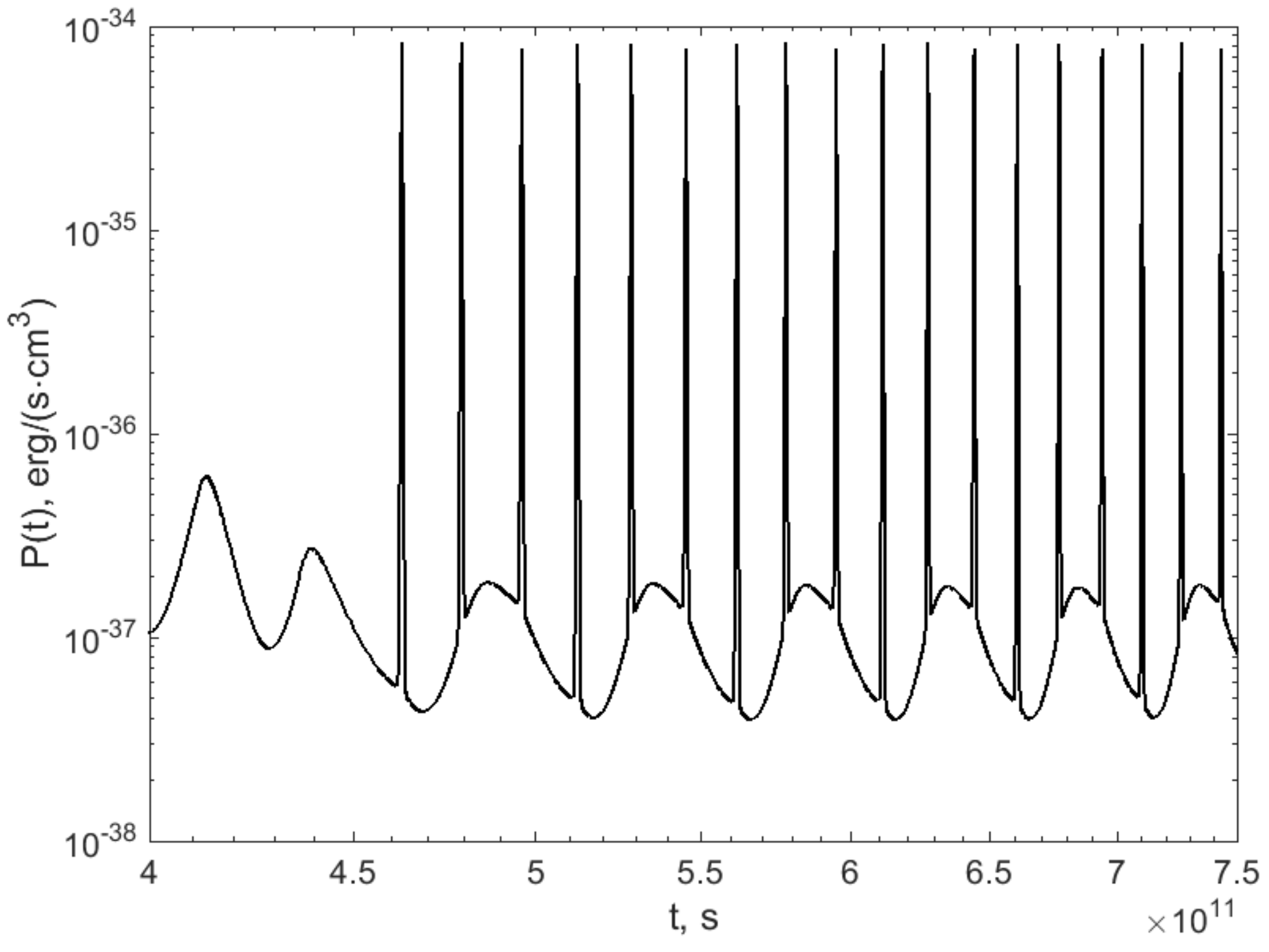}
\caption{Power losses of the pair beam, Eq. (\ref{powerloss}), for case 4. }
\label{Pcase4}
\end{figure}

\begin{figure}
\includegraphics[width = \columnwidth]{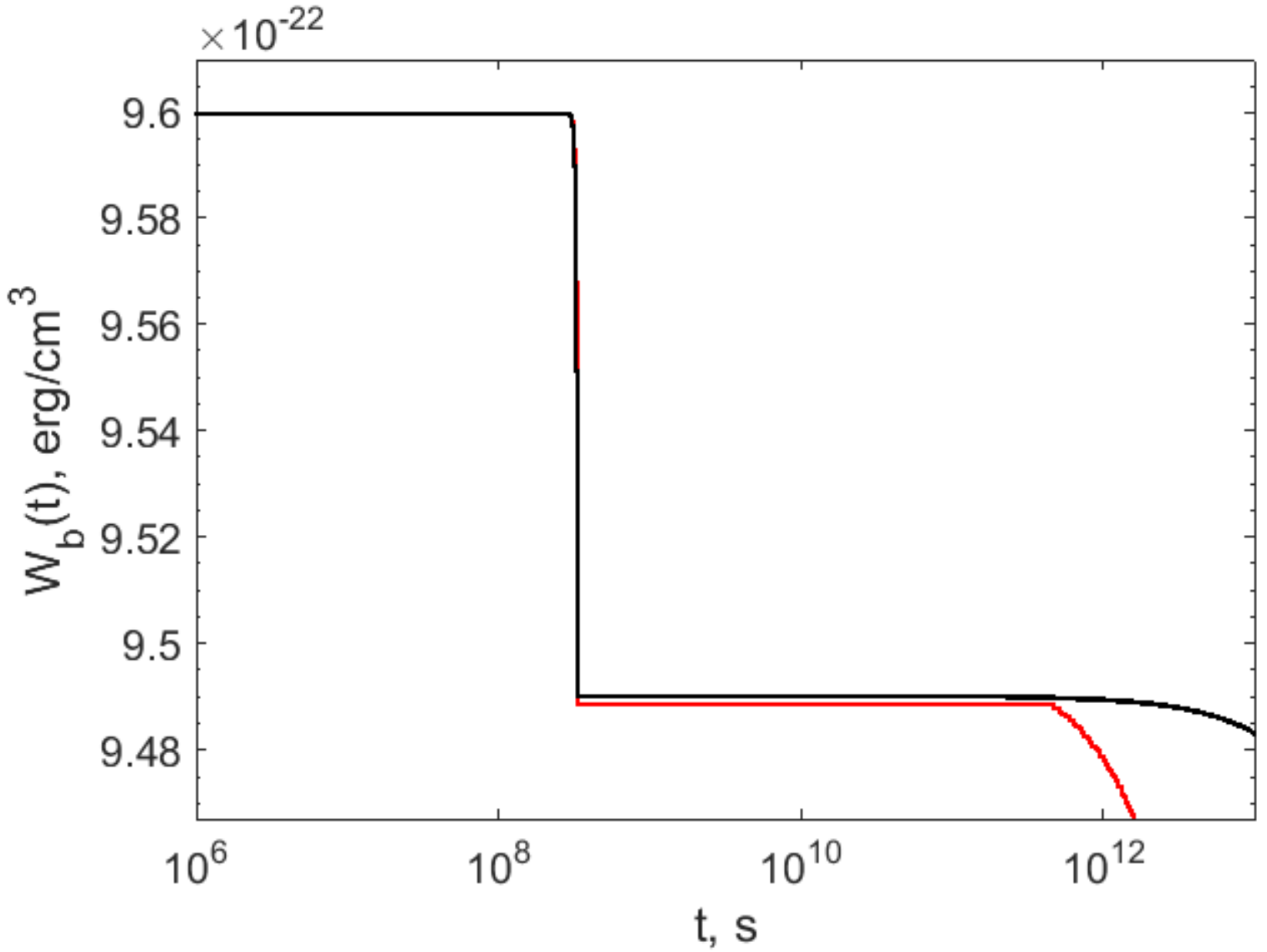}
\caption{Evolution of the beam energy density for cases 1(black) and 4(red).}
\label{Wbt}
\end{figure}

% \begin{table}
% \caption{Relaxation time with collisions} % title of Table
% \centering % used for centering table
% \begin{tabular}{|c|c|} % centered columns (4 columns)
% \hline %inserts double horizontal lines
% Case & $\tau_{rel}$, s  \\ [0.5ex] 
% \hline %horizontal line
% 4 & $2.7\cdot10^{14}$  \\
% \hline
% 5 & $2.2\cdot10^{14}$   \\
% \hline
% 6 & $8.3\cdot10^{13}$  \\
% \hline
% \end{tabular}
% \label{Table4} % is used to refer this table in the text
% \end{table}

\section{Summary and conclusions}\label{summary}
We revisited the effect of NL damping on the relaxation time of the blazar-induced pair beam, $\tau_{rel}$. The relaxation time of the pair beams is given by the wave growth rate and the wave intensity in the resonant band of the spectrum, and so one needs to know the saturation level of the wave intensity and its stability, as the relaxation time is far longer than the time needed to reach saturation. NL damping is an important factor in the nonlinear evolution of the intensity of the electrostatic plasma waves that the pair beams drive. During the quasi-stationary phase, its efficiency primarily depends on the wave intensity at small $k$ (where it is much higher than anywhere else in the spectrum), which it provides itself by transferring wave energy from the resonant band of the spectrum, implying a high level of nonlinearity. We investigate this problem numerically using a simplified 2D model which we demonstrate to be an acceptable approximation of the real 3D physics. We also consider a more realistic growth rate of the electrostatic instability than is found in, e.g., \citet{Chang14}. 

In an earlier study \citep{Vafin18} we used a simplified description of NL damping and found it weaker than the modulation instability and both permitting a saturation intensity, for which the relaxation time of the pair beam is shorter than the IC time ($\tau_{IC}=2.5\cdot10^{13}$ s). This new study is based on a detailed calculation of the intensity spectrum and its evolution, which should provide an accurate estimate of the relaxation time of the beam.

\citet{Puchwein12} argued that the dissipation of pair-beam energy may be an efficient means of IGM heating. Since the NL damping rate exponentially depends on the plasma temperature, we performed several simulations to explore the effect of the IGM temperature on $\tau_{rel}$. At the average IGM temperature of 1 eV, the relaxation time of the pair beam is much longer than at $T_\mathrm{IGM}=8$~eV.

We find that quasi-perpendicular waves ($k_\parallel c/\omega_{p,e}\simeq 1$ and $k_\perp c/\omega_{p,e}\gg 1$) display a complicated and time-variable spectrum whose intensity is typically too low to impose significant beam dissipation. In the spectral band where the growth rate is highest ($k_\parallel c/\omega_{p,e}\simeq 1$ and $k_\perp c/\omega_{p,e}\approx 1$), the wave intensity is completely drained by NL damping driven by very high wave intensity at small $k$. Taken at face value, these results suggest that the beam dissipates a negligible portion of its energy during the IC cooling time. However, if other dissipation mechanisms are present that can efficiently dissipate the non-resonant waves at small $k$ that are responsible for a high NL damping rate, then the relaxation time of the beam would be significantly reduced. Among the possible candidates are the modulation instability and the collisional damping. In the current work, we considered only the latter. 

Since the effect of collisions becomes important at small plasma temperatures, we performed several calculations with $T_\mathrm{IGM}<1$ eV including collisional damping. We found that the relaxation time is indeed considerably shorter than without collisions. This is surprising in view of the collision rate being about four orders of magnitude smaller than the peak growth rate of the electrostatic waves, and it reflects the delayed build-up of wave intensity at small $k$ by NL damping. Formally, even for $T_\mathrm{IGM}=0.3$ eV, the relaxation time is still longer than the IC time. The large impact of a seemingly sub-dominant process indicates that the beam dissipation rate is extremely sensitive to the dissipation processes under consideration and to the accuracy of their treatment. The latter is best seen in the difference in estimated relaxation time between approximating NL damping by a simple decay term (as in \citet{Vafin18}) and explicitly following the evolution of the wave spectrum, as done here. Thus, any other damping process (e.g., the modulation instability) will further modify the dissipation rate, and a full spectral treatment of all damping and cascading processes is needed to reliably calculate the beam dissipation rate, but that is a topic of future investigations.

The effect of collisional damping in the beam dissipation process is substantially small in the sense that the relaxation time is much larger than the IC scattering time even in their presence. Moreover, it is inversely dependent on the IGM temperature. In principle, for a sufficiently cold plasma, collisions could play a major role in beam energy dissipation, despite it being one of the slowest processes. Other processes like wave-wave scattering or wave-particle scattering produce higher rates thereby dominating the effect of particle collisions \citep{Chang14}. Other commonly ignored processes may play a role as well, and so an accurate estimate of the dissipation rate, and consequently, the relaxation time requires one to consider all possible damping mechanisms. Since collisional damping is not a predominant factor in the beam dissipation process, a direct constraint on the IGM temperature from its effect appears to be out of reach.

There is observational evidence for beam dissipation arising from the absence of GeV haloes around the jets of misaligned AGN \citep{Broderick18}. This paper demonstrates that theoretically calculating the beam dissipation rate is very challenging on account of the inherent non-linearity, and simple estimates may be misleading. It is conceivable that beam dissipation is efficient only for beams of a certain density, which translates to a maximum distance, $L_\mathrm{max}$, from the AGN that depends on its multi-TeV gamma-ray flux. We are not yet in the position to reliably calculate that distance and hence to estimate which part of the cascade would be quenched (at $L<L_\mathrm{max}$) and what spectrum the observable cascade produced at $L>L_\mathrm{max}$ would have. 

%%%%%%%%%%%%%%%%%%%%%%%%%%%%%%%%%%%%%%%%%%%%%%%%%%%%%%%%%%%%%%%%%%%%%%%%%%%%%%%%%%%%%%%%%%%%%%%%%%%%%%%%%%%%%%%%%%%%%%%%%%%%%%%%%%%%
\acknowledgments

%%%%%%%%%%%%%%%%%%%%%%%%%%%%%%%%%%%%%%%%%%%%%%%%%%%%%%%%%%%%%%%%%%%%%%%%%%%%%%%%%%%%%%%%%%%%%%%%%%%%%%%%%%%%%%%%%%%%%%%%%%%%%%%%%%%%%%
\appendix
\section{Approximate expression for $\omega_{NL}({\bf \MakeLowercase{k}})$ in 3D model}\label{appwnl}

Eq. (\ref{nlrate}) for $\omega_{NL}({\bf k})$ is a three dimensional integral that can be written as 
\be 
\omega_{NL}(k_\parallel,k_\perp)= {3(2\pi)^{1/2}\over 64 n_e m_e u_i} \int k_\perp'dk_\perp' d\phi' dk_\parallel' I W(k_\parallel',k_\perp') 
{  k'^2-k^2  \over (k' k)^2},
\label{app1}
\ee 
where
\be 
I= {( k_\parallel'k_\parallel + k_\perp'k_\perp \cos\phi' )^2 \over \l[ (k_\parallel'-k_\parallel)^2 + k_\perp'^2+k_\perp^2 - 2k_\perp'k_\perp\cos\phi' \r]^{1/2}} \exp\l[ - a \l( {c\over\omega_{p,e}} {k'^2-k^2 \over \l[ (k_\parallel'-k_\parallel)^2 + k_\perp'^2+k_\perp^2 - 2k_\perp'k_\perp\cos\phi' \r]^{1/2} } \r)^2\r].
\label{app2}
\ee

It is easy to see that for 
\be
k_\perp'\gg k_\perp \vee k_\perp'\ll k_\perp \vee |k_\parallel'-k_\parallel|\gg \sqrt{2k_\perp' k_\perp}
\label{app4}
\ee
one can neglect $\cos\phi'$ under the square roots in Eq. (\ref{app2}) yielding 
\be 
\int_0^{2\pi} I d\phi'= 
\pi{ 2(k_\parallel'k_\parallel)^2 + (k_\perp'k_\perp)^2   \over \l[ (k_\parallel'-k_\parallel)^2 + k_\perp'^2+k_\perp^2 \r]^{1/2}} \exp\l[ - a \l( {c\over\omega_{p,e}} {k'^2-k^2 \over \l[ (k_\parallel'-k_\parallel)^2 + k_\perp'^2+k_\perp^2  \r]^{1/2} } \r)^2\r],
\label{app3}
\ee
while in the region $k_\perp'\sim k_\perp \wedge |k_\parallel' -k_\parallel|\lesssim \sqrt{2}k_\perp$ 
\be 
\int_0^{2\pi}Id\phi'\approx \int_0^{2\pi}d\phi' 
{( k_\parallel'k_\parallel + k_\perp^2 \cos\phi' )^2 \over \l[ (k_\parallel'-k_\parallel)^2 + 2k_\perp^2(1 - \cos\phi') \r]^{1/2}} \exp\l[ - a \l( {c\over\omega_{p,e}} {k_\parallel'^2-k_\parallel^2 \over \l[ (k_\parallel'-k_\parallel)^2 + 2k_\perp^2(1 - \cos\phi') \r]^{1/2} } \r)^2\r]. 
\label{app5}
\ee
If $\phi'$ is not very close to 0 so that $|1-\cos\phi'|$ is order of 1 ($0.3\pi\lesssim\phi'\lesssim1.7\pi$, $\cos\phi\lesssim0.5$) then the integral (\ref{app5}) can be again estimated neglecting $\cos\phi'$ under the square roots with the result which is order of Eq. (\ref{app3}) at $k_\perp'\sim k_\perp$:
\begin{multline} 
\int_{0.3\pi}^{1.7\pi} Id\phi' \sim \int_0^{2\pi} d\phi'
{( k_\parallel'k_\parallel + k_\perp^2 \cos\phi' )^2 \over \l[ (k_\parallel'-k_\parallel)^2 + 2k_\perp^2 \r]^{1/2}} \exp\l[ - a \l( {c\over\omega_{p,e}} {k_\parallel'^2-k_\parallel^2 \over \l[ (k_\parallel'-k_\parallel)^2 + 2k_\perp^2 \r]^{1/2} } \r)^2\r]= \\
\pi{ 2(k_\parallel'k_\parallel)^2 + (k_\perp)^4   \over \l[ (k_\parallel'-k_\parallel)^2 + 2k_\perp^2 \r]^{1/2}} \exp\l[ - a \l( {c\over\omega_{p,e}} {k_\parallel'^2-k_\parallel^2 \over \l[ (k_\parallel'-k_\parallel)^2 + 2k_\perp^2  \r]^{1/2} } \r)^2\r]. 
\label{app6}
\end{multline}
At the same time, the contribution to the integral (\ref{app5}) from the angles satisfying $|1-\cos\phi'|\ll(k_\parallel'-k_\parallel)^2/(2k_\perp^2)$ ($\phi'\lesssim \Delta\ll \sqrt{2}|k_\parallel'-k_\parallel|/k_\perp$) is 
\be 
\int_{2\pi-\Delta}^{0+\Delta}Id\phi' \approx \Delta {( k_\parallel'k_\parallel + k_\perp^2  )^2 \over |k_\parallel'-k_\parallel|} \exp\l[ - a \l( {c\over\omega_{p,e}} {k_\parallel'^2-k_\parallel^2 \over |k_\parallel'-k_\parallel | } \r)^2\r]
\label{app7}
\ee
which is much smaller compared to Eq. (\ref{app6}). Thus, Eq. (\ref{app3}) can be used in the whole $(k_\perp',k_\parallel')$-plane to approximately calculate the $\phi'$-integral in the expression for $\omega_{NL}({\bf k})$. 

\section{Approximate expression for $\omega_{NL}({\bf \MakeLowercase{k}})$ in 2D model}\label{appwnl1d}
In the 2D model
\be 
\omega_{NL}(k_\perp)= {3(2\pi)^{1/2}\over 64 n_e m_e u_i} \int k_\perp'dk_\perp' d\phi' G W(k_\perp') 
{  k_\perp'^2-k_\perp^2  \over (k_\perp' k_\perp)^2},
\label{appB1}
\ee 
where
\be 
G= {(  k_\perp'k_\perp \cos\phi' )^2 \over \l[ k_\perp'^2+k_\perp^2 - 2k_\perp'k_\perp\cos\phi' \r]^{1/2}} \exp\l[ - a \l( {c\over\omega_{p,e}} {k_\perp'^2-k_\perp^2 \over \l[  k_\perp'^2+k_\perp^2 - 2k_\perp'k_\perp\cos\phi' \r]^{1/2} } \r)^2\r].
\label{appB2}
\ee
The same proceeding as in Appendix \ref{appwnl} yields
\be 
\int_0^{2\pi} G d\phi' \approx
\pi{  (k_\perp'k_\perp)^2   \over \l(  k_\perp'^2+k_\perp^2 \r)^{1/2}} \exp\l[ - a \l( {c\over\omega_{p,e}} {k_\perp'^2-k_\perp^2 \over \l( k_\perp'^2+k_\perp^2  \r)^{1/2} } \r)^2\r].
\label{appB3}
\ee

\section{Applicability of 2D model}\label{applicability2D}

Here, we provide arguments for the applicability of the 2D model to study Eq. (\ref{specteq}). The evolution of the spectrum in 3D as well as 2D exhibits two characteristic stages: linear and non-linear. During the linear stage, the non-linear Landau damping is negligible compared to the growth rate, because the spectral energy $W({\bf k})$ is too small. The system turns to the non-linear regime when the NL damping rate becomes comparable to the maximum growth rate of the instability. Let us denote the time when it takes place by $t=t_{NL}$. We want to demonstrate now that the 2D model reflects the main features of the original 3D one during the both linear and non-linear regimes.

\subsection{Linear regime}

During the linear regime $t\ll t_{NL}$, the wave energy growth exponentially $\propto\exp(2\max(\omega_i)t)$ in both 3D and 2D model where the maximum growth rate $\max(\omega_i)$ in 2D is the same as in 3D. 

At the transition time $t\approx t_{NL}$, the NL damping starts to balance the instability growth. We can estimate the electric filed energy at this time. Let us consider a time $t$ close enough to $t_{NL}$ and such that the electric filed energy is still concentrated near the wave vector with the maximum growth rate (at $t\gg t_{NL}$, the most waves will be scattered to small wave vectors where the growth rate is zero). In the 3D model, the most unstable wave number is located near $k\sim\omega_{p,e}/c$ ($k_\perp\sim\omega_{p,e}/c$ in 2D). In the vicinity of this wave vector, we can neglect the exponential function in Eq. (\ref{nlrate}) (Eq. (\ref{nlrate1d}) in 2D) because the factor $a\ll1$. Additionally, we replace $({\bf k}'{\bf k})^2/(k'^2k^2)$ in Eq. (\ref{nlrate}) by its angle-averaged value $1/3$ ($({\bf k}_\perp'{\bf k}_\perp)^2/(k_\perp'^2k_\perp^2)$ in Eq. (\ref{nlrate1d}) is replaced by $1/2$). Then, since the width of growth rate in the parallel direction $\Delta k_\parallel$ in 3D is extremely narrow ($\Delta k_\parallel\ll \omega_{p,e}/c$), we also adopt ${\bf k}_\parallel'\approx {\bf k}_\parallel$  and ${\bf k}_\parallel'\approx -{\bf k}_\parallel$ for positive and negative $k_\parallel'$, accordingly. Finally, the NL damping rate in 3D and 2D cases becomes, respectively, 
\begin{multline}
\omega_{NL}^{3D}(k\approx\omega_{p,e}/c)\approx {(2\pi)^{1/2} \over 64 n_e m_e u_i} \int d^2k_\perp' W_{3D}({\bf k}')\Delta k_\parallel
\l( {k_\perp'^2-k_\perp^2 \over |{\bf k}_\perp'-{\bf k}_\perp|} + {k_\perp'^2-k_\perp^2 \over |{\bf k}_\perp'-{\bf k}_\perp-2{\bf k}_\parallel|} \r)=\\
 {\pi^{1/2} \over 64 n_e m_e u_i} \int d^2k_\perp' W_{3D}({\bf k}')\Delta k_\parallel {k_\perp'^2-k_\perp^2 \over |{\bf k}_\perp'-{\bf k}_\perp|}
\l( 1 + { |{\bf k}_\perp'-{\bf k}_\perp| \over |{\bf k}_\perp'-{\bf k}_\perp-2{\bf k}_\parallel|} \r),
\label{appc1}
\end{multline}
and
\be 
\omega_{NL}^{2D}(k_\perp\approx\omega_{p,e}/c)\approx {3(2\pi)^{1/2} \over 128 n_e m_e u_i} \int d^2k_{\perp}' W_{2D}(k_\perp')
{k_\perp'^2-k_\perp^2 \over |{\bf k}_\perp'-{\bf k}_\perp|}.
\label{appc2}
\ee
Since the term $ |{\bf k}_\perp'-{\bf k}_\perp| / |{\bf k}_\perp'-{\bf k}_\perp-2{\bf k}_\parallel|<1$ and $W_{2D}(k_\perp)\approx W_{3D}({\bf k})\Delta k_\parallel$, Eq. (\ref{appc2}) differs from Eq. (\ref{appc1}) at most by factor 3/2. Let us continue to work with Eq. (\ref{appc2}) to derive the electric field energy at $t\approx t_{NL}$. Using the same technique as in Appendix \ref{appwnl} and \ref{appwnl1d}, Eq. (\ref{appc2}) yields 
\be 
\omega_{NL}^{2D}(k_\perp\approx\omega_{p,e}/c)\approx {3(2\pi)^{1/2} \over 128 n_e m_e u_i} \int d^2k_{\perp}' W_{2D}(k_\perp')
{k_\perp'^2-k_\perp^2 \over \sqrt{k_\perp'^2+k_\perp^2}}
\label{appc3}
\ee
where the integral is calculated in the vicinity the most unstable mode $k_\perp\approx \omega_{p,e}/c$. Then, $k_\perp'$ is close to $k_\perp$ and Eq. (\ref{appc3}) becomes
\be 
\omega_{NL}^{2D}(k_\perp\approx\omega_{p,e}/c)\approx {3\pi^{1/2} \over 64 n_e m_e u_i} \int d^2k_{\perp}' W_{2D}(k_\perp') (k_\perp'-k_\perp)\approx 
{3\pi^{1/2} \over 64 n_e m_e u_i} W_{E}^{tot} \Delta k_\perp
\label{appc4}
\ee
where $W_E^{tot}$ is the total electric field energy density and $\Delta k_\perp$ in the characteristic spectral width in the perpendicular direction at $t\approx t_{NL}$. At the transition point, $\omega_{NL}\approx\max(\omega_i)$ and 
\be 
W_E^{tot}= {64 n_e m_e u_i \max(\omega_i) \over 3\pi^{1/2} \Delta k_\perp}.
\label{appc5}
\ee
Since Eq. \ref{appc1} differs from Eq. \ref{appc2} only by factor $2/3$, $W_E^{tot}$ in the 3D model will be of the same order as Eq. (\ref{appc5}) derived for the 2D one. Furthermore, our numerical results (see section \ref{results}) show that $W_E^{tot}$ constitutes about $1$ \% of the the initial beam energy density. Therefore, an additional factor $2/3$ in the 3D model won't principally affect this result.  Therefore, the the wave scattering in the ${\bf k}_\parallel$-direction absent in the 2D model is not crucial for the beam energy loses during the linear phase and can be, indeed, neglected.

\subsection{Non-linear regime}

Now, we turn to the non-linear spectral evolution at times $t\gg t_{NL}$. After the NL damping becomes significant and scatters the most wave energy to lower wave numbers, the further energy growth of the electric field is regulated by the spectrum at large wave numbers ($k_\perp c/\omega_{p,e}\gg1$) where both LL and NL damping rates are smaller than the instability growth rate \citep{Chang14}. However, the spectrum at $k_\perp c/\omega_{p,e}\approx 1$ also contributes to the electric field growth at $t\gg t_{NL}$ if particle collisions are present as we found in our 2D model (see section \ref{collisions}). There should exist a similar effect in the 3D model near the region of the maximum growth rate $kc/\omega_{p,e}\approx1$. Thus, there can be two resonant groups of waves driving the instability at $t\gg t_{NL}$. At the same time, there is also a group of non-resonant waves at small wave vectors $k\sim0$ ($k_\perp\sim0$ in 2D). We sketched these three different parts of the spectrum schematically in Fig. \ref{SpectrumSketch}. We note that the position of the red circle (corresponding to resonant waves growing when the collisional damping is present) is slightly different for the 3D and 2D models. The reason simply because the maximum growth rate in our 2D model occurs at $k_\perp c/\omega_{p,e}=1$, while in the 3D one it is $k_\perp c/\omega_{p,e}<1$. Another difference between 2D and 3D is that the spectrum in 3D is actually symmetric for negative $k_\parallel$, the 2D model treats the spectrum evolution only in the plane $k_\parallel c/\omega_{p,e}=1$. We will show now that the 2D and 3D models provide close values  of the NL damping rate in the resonant regions (blue and red circles in Fig. \ref{SpectrumSketch}) which are decisive for the beam energy losses, because only the resonant waves can interact with the beam. 

\subsubsection{Resonant waves at large wave vectors}\label{resonantwaves}

Let us first consider the resonant region marked by the blue circle in Fig. \ref{SpectrumSketch}. It is rather straightforward that the regions located at much smaller wave vectors (green and and red circles) in 2D and 3D provide close contributions to the NL damping, since the difference between the positions of green and red circles in the 2D and 3D models is negligible for the NL damping rate at large wave vectors. To demonstrate this explicitly, we can neglect $|{\bf k}'|\ll|{\bf k}|$ and approximately replace $({\bf k}'{\bf k})^2/(k'^2k^2)$ in Eq. (\ref{nlrate}) by its angle-averaged value $1/3$. Then, Eq. (\ref{nlrate}) becomes

\begin{equation} 
\omega_{NL}^{3D}(k_\perp c/\omega_{p,e}\gg1; |k_\parallel'|c/\omega_{p,e}\sim1, k'c/\omega_{p,e}\sim0)= -{(2\pi)^{1/2}\over 64 n_e m_e u_i}k\exp\l[ - a \l( {kc\over\omega_{p,e}}  \r)^2\r]  W^E_{tot}(|k_\parallel'|c/\omega_{p,e}\sim1, k'c/\omega_{p,e}\sim0) ,
\label{appc6}
\end{equation}
A similar derivation in 2D leads to 
\begin{equation} 
\omega_{NL}^{2D}(k_\perp c/\omega_{p,e}\gg1; k_\perp'c/\omega_{p,e}\sim1, k_\perp'c/\omega_{p,e}\sim0)= -{3(2\pi)^{1/2}\over 128 n_e m_e u_i} k_\perp \exp\l[ - a \l( {k_\perp c\over\omega_{p,e}}\r)^2 \r]W^E_{tot}(k_\perp'c/\omega_{p,e}\sim1, k_\perp'c/\omega_{p,e}\sim0).
\label{appc7}
\end{equation}
The total energies in Eqs. (\ref{appc6}) and (\ref{appc7}) are expected to be comparable, since we found above that the electric field gains close energies in the 2D and 3D model during the linear stage. Then the difference between Eq. (\ref{appc6}) and (\ref{appc7}) is only factor $2/3$. Thus, our 2D model provides a reasonable form of spectrum at small wave vectors that does not considerable modify the NL damping in the resonant region at large wave numbers. To finish the analysis of the NL damping in the resonant region at large wave vectors, we need to consider the contribution into NL damping rate from another resonant region at large wave vectors but with negative $k_\parallel<0$. We note that this region is absent in the 2D model, since we study the problem in the plane $k_\parallel c/\omega_{p,e}=1$. We will show below that this contribution in 3D is negligible and our 2D model is justified in this respect. Setting ${\bf k}_\perp'\approx {\bf k}_\perp$ and $k_\parallel'\approx-k_\parallel$, we arrive at 
\begin{equation} 
\omega_{NL}^{3D}(k_\perp c/\omega_{p,e}\gg1; k_\parallel'c/\omega_{p,e}\sim-1, k_\perp'c/\omega_{p,e}\gg1)\approx- {3(2\pi)^{1/2}\over 64 n_e m_e u_i}\Delta k_\parallel  W^E_{tot}(k_\parallel'c/\omega_{p,e}\sim-1, k_\perp'c/\omega_{p,e}\gg1) ,
\label{appc8}
\end{equation}
Except from the region where the exponential factor in Eq. (\ref{appc6}) is small enough, Eq. (\ref{appc6}) gives a much larger contribution into the NL damping rate than Eq. (\ref{appc8}), as $\Delta k_\parallel\ll k$ and the most electric field energy  energy is concentrated at small wave vectors. Summarizing the above derivation, the NL damping rate at large $kc/\omega_{p,e}\gg1$ in 3D is close to that in 2D.         

\subsubsection{Resonant waves at small wave vectors due to collisional effect}

\begin{figure}
\centering
\includegraphics[width = 87.5mm]{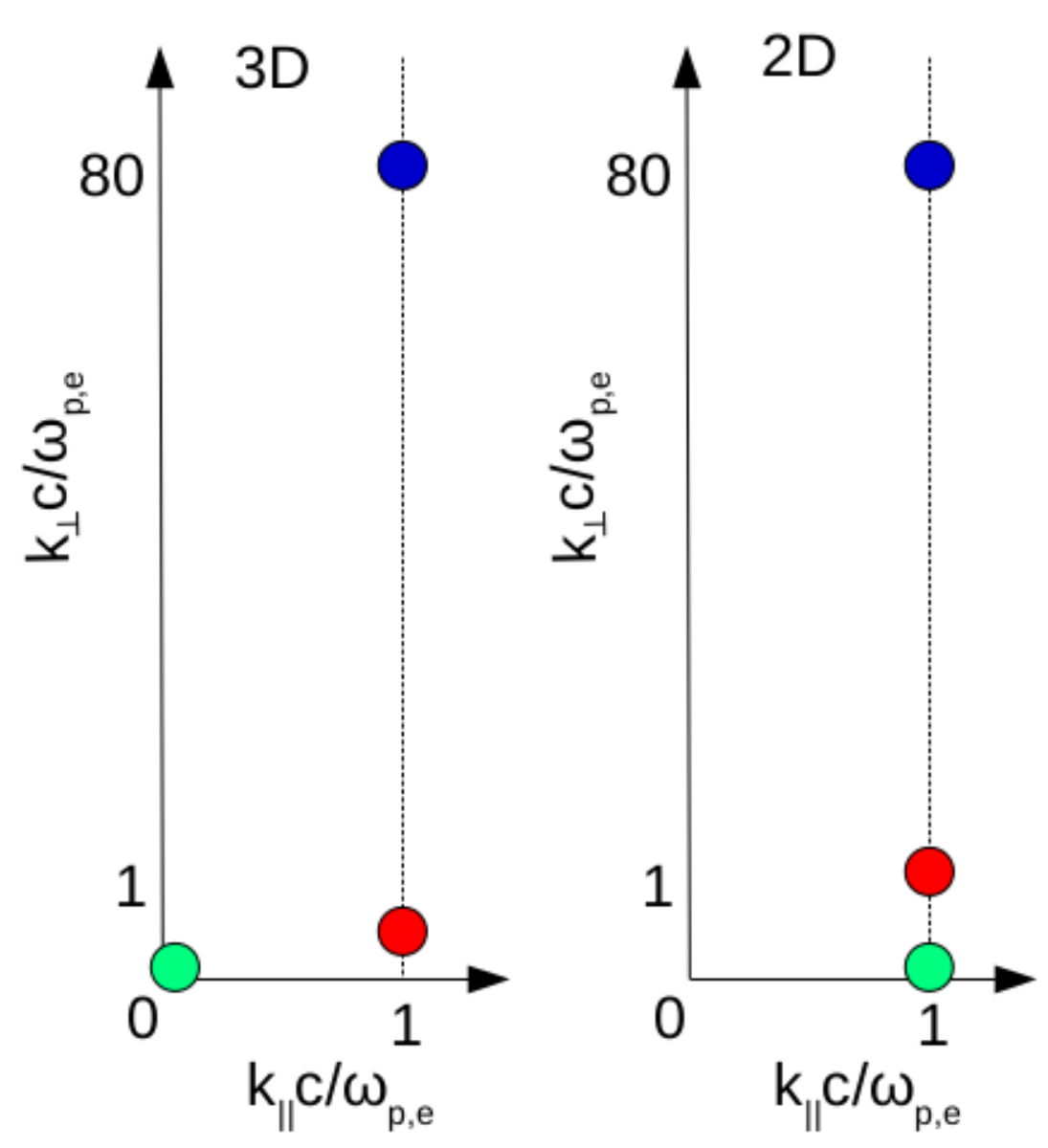}
\caption{Schematic picture of the electric field spectrum in the 3D (left panel) and 2D (right panel) models. Green: non-resonant waves. Blue: resonant waves at large $k_\perp c/\omega_{p,e}$ where the NL damping is suppressed. Red: resonant waves that can also exist when particle collisions are present.}
\label{SpectrumSketch}
\end{figure}

Now, let us consider another resonant region that can appear at $t\gg t_{NL}$ due to collisional damping of the electric field energy (see section \ref{collisions}) and shown by the red circle in Fig. \ref{SpectrumSketch}. First, we note that the position of the red circle that we expect in 3D is not far from that one in 2D, since the maximum growth rate in 3D occurs at $k_\perp c/\omega_{p,e}\approx 0.6$, while in 2D it is $k_\perp c/\omega_{p,e}\approx1$. Second, similar to section \ref{resonantwaves} it can be demonstrated  that the integral contribution into the NL damping rate from resonant waves at large wave vectors (blue circle) differs slightly for the 2D and 3D models, because the difference in the wave vectors between those groups of waves is large. Finally, it is left to compare the NL damping rate due to the interaction with the non-resonant waves (green circle) (it can be proved in a similar manner as in section \ref{resonantwaves} that the contribution from the symmetric region at $k_\parallel c/\omega_{p,e}\approx-1$ in 3D is smaller than from the non-resonant waves at $k\approx0$ in 3D). To do this, we can utilize similar simplifications as we used above to derive Eq. (\ref{appc1}). Again, we neglect the exponential factor and replace $({\bf k}'{\bf k})^2/(k'^2k^2)$ in Eq. (\ref{nlrate}) by its angle-averaged value $1/3$    
\begin{equation} 
\omega_{NL}^{3D}(k_\perp c/\omega_{p,e}\sim1; k'c/\omega_{p,e}\sim0)= -{(2\pi)^{1/2}\over 64 n_e m_e u_i}k  W^E_{tot}(k'c/\omega_{p,e}\sim0) ,
\label{appc9}
\end{equation}
A similar derivation in 2D leads to 
\begin{equation} 
\omega_{NL}^{2D}(k_\perp c/\omega_{p,e}\sim1; k_\perp'c/\omega_{p,e}\sim0)= -{3(2\pi)^{1/2}\over 128 n_e m_e u_i} k_\perp W^E_{tot}( k_\perp'c/\omega_{p,e}\sim0).
\label{appc10}
\end{equation}

Thus, the NL damping rates in 2D and 3D are close to each other also for resonant waves at small wave vectors. This completes the analysis of applicability of the 2D model.

\section{Numerical comparison of 3D and 2D models}\label{comparison3D2D}

\begin{figure}
    \centering
    \begin{tabular}{cc@{}}
	\subfloat[]{\includegraphics[width = 0.47\columnwidth]{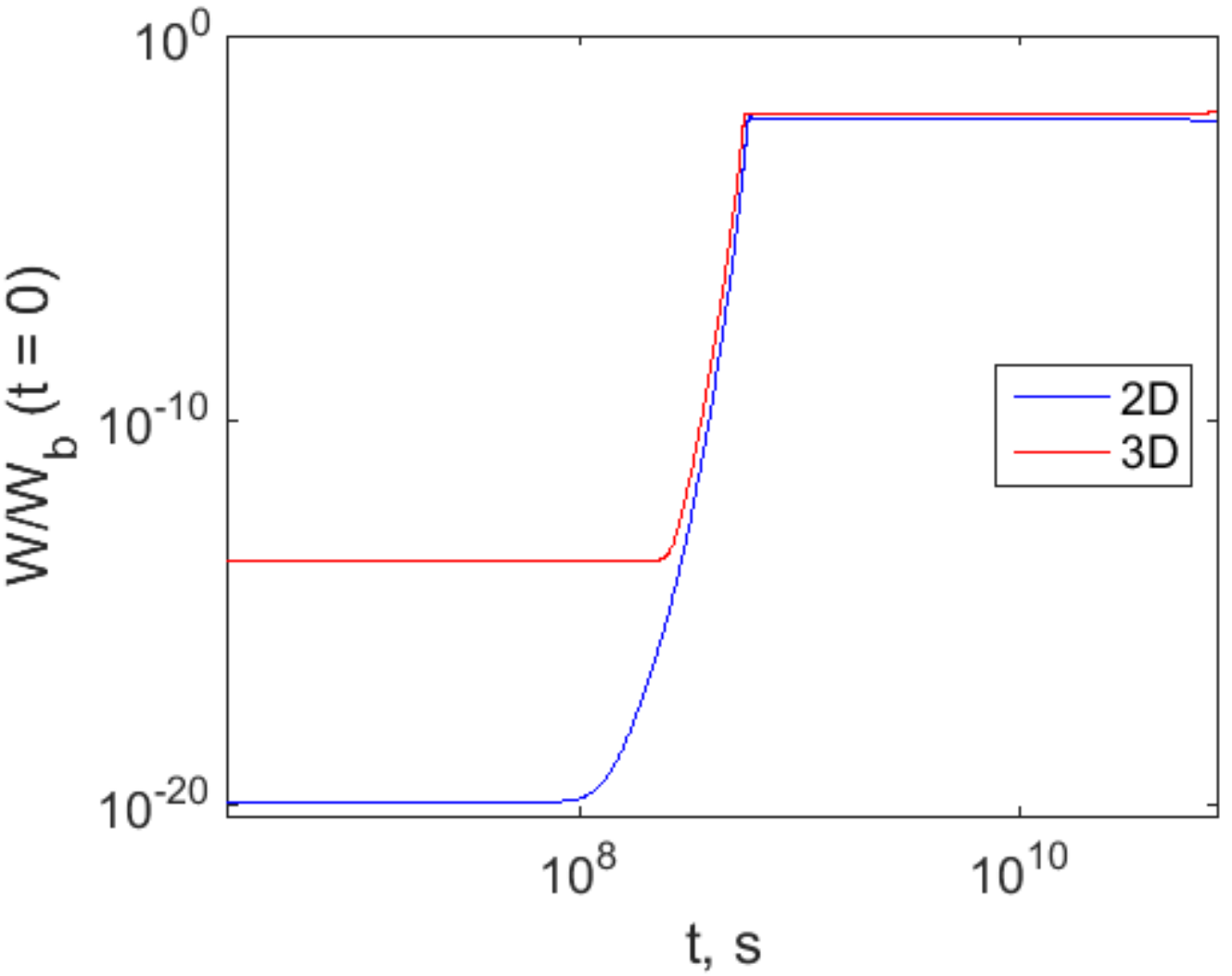}}
    \hspace{0.5cm}
	\subfloat[]{\includegraphics[width = 0.47\columnwidth]{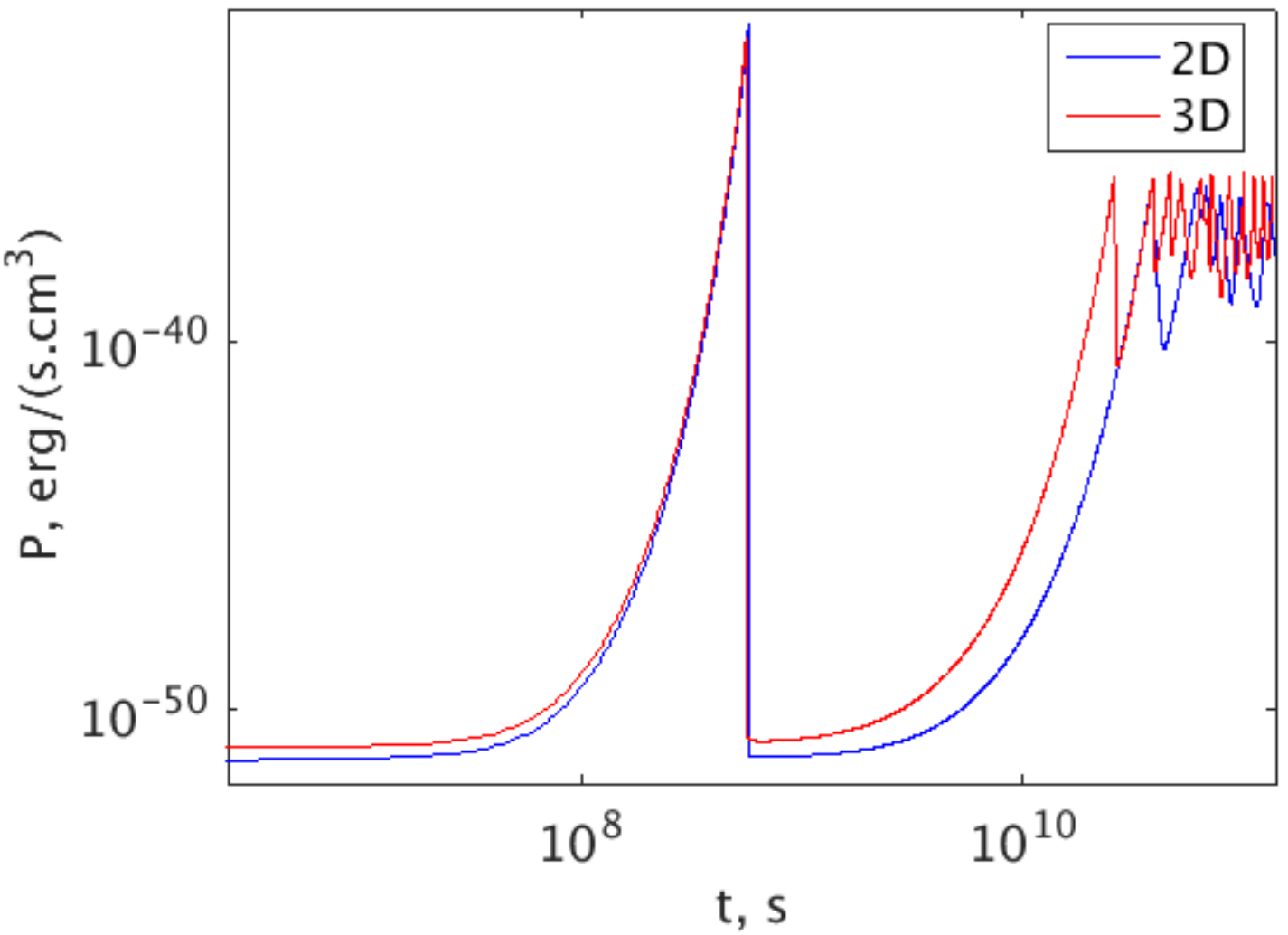}}
    \end{tabular}   
    \caption{A comparison of (a) total electric field energy density, and (b) power losses as a function of time between 2D (blue curve) and 3D (red curve) for case 3.}
    \label{2d3dcomp}
\end{figure}

To compare the 3D and 2D models, we chose the parameters of case 3 in Table \ref{Table1}, since the simulation region is the smallest ($k_\perp c/\omega_{p,e}<35$, $k_\parallel c/\omega_{p,e}\lesssim1$) among all cases. Although the spatial resolution in 3D has been compromised owing to computational costs, the calculations should provide a reasonable estimate of the comparability of the 2D and the 3D models in the linear and early nonlinear stages of evolution. The region of instability ($\delta k_{||}c/\omega_{p,e} = 10^{-6}$ at $k_{||}c/\omega_{p,e} = 1$) is resolved with 7 linearly spaced grid points, and we employ 60 logarithmically spaced grid points between $k_{||}c/\omega_{p,e} = 10^{-7}$ and $k_{||}c/\omega_{p,e} = 1$ to cover the region to which NL damping transfers wave energy. In the transverse direction, we use 40 logarithmically spaced grid points to resolve the spectrum at $k_{\perp}c/\omega_{p,e} \leq 1$, where the growth rate remains essentially independent of $k_\perp$. In the band $k_{\perp}c/\omega_{p,e} \geq 1$, we have 110 linearly spaced grid points extending up to $k_{\perp}c/\omega_{p,e} = 35$, beyond which linear Landau damping prohibits wave growth (cf. Fig.~\ref{WiWLL}).

Fig. \ref{2d3dcomp} (a) illustrates the total electric field energy density normalized to the beam energy density for the 2D and 3D scenarios. It is clearly seen that the energy evolution in 2D case replicates the 3D case except for the noise level. In 2D, the noise is given by $k_BT_\mathrm{IGM}\Delta k_\parallel$ to account for the energy concentrated in the actual region of instability ($\Delta k_\parallel c/\omega_{p,e}= 10^{-6} $), whereas for 3D it is simply given by $k_BT_\mathrm{IGM}$. Nevertheless, the electric field energy turns to quasi-saturation at the value much smaller than the initial beam energy density.  A similar comparison of the power losses is shown in Fig. \ref{2d3dcomp} (b).   The power losses in 3D were calculated similar to Eq. (\ref{powerloss}):
\be 
P(t)=8\pi\,\int\,\omega_i W({\bf k})k_\perp dk_\perp dk_\parallel.
\label{powerloss3D}
\ee

Although the nonlinear effects set in slightly earlier in the 3D scenario, the power losses in the 2D model are in good agreement with the 3D one. Finally, we conclude that our 2D model can be reasonably used to study Eq. (\ref{specteq}).


\begin{thebibliography}{}

\bibitem[Ackermann et al.(2012)]{Ackermann12}Ackermann, M., Ajello, M., Allafort, A., et al.\ 2012, Sci., 338, 1190
\bibitem[Aharonian et al.(1994)]{1994ApJ...423L...5A} Aharonian, F.~A., Coppi, P.~S., \& Voelk, H.~J.\ 1994, \apjl, 423, L5 
\bibitem[Alexandrov et al.(1984)]{Alexandrov84}Alexandrov, A. F., Bogdankevich, L. S., \& Rukhadze, A. A.\ 1984, Principles of Plasma Electrodynamics (Springer-Verlag Berlin Heidelberg)
\bibitem[Baikov(1977)]{Baikov77}Baikov, I. S.\ 1977, Physica, 92C, 267
\bibitem[Breizman et al.(1972)]{Breizman72}Breizman, B. N., Ryutov, D. D., \& Chebotaev, P. Z.\ 1972, JETP, 35, 741
\bibitem[Breizman(1990)]{Breizman90}Breizman, B. N.\ 1990, RvPP, 15, 61
\bibitem[Broderick et al.(2012)]{Broderick12}Broderick, A. E., Chang, P., \& Pfrommer, C.\ 2012, ApJ, 752, 22
\bibitem[Broderick et al.(2016)]{Broderick16}Broderick, A. E., Tiede P., Shalaby, M.\ 2016, ApJ, 832, 109
\bibitem[Broderick et al.(2018)]{Broderick18}Broderick, A. E., Tiede, P., Chang, P., et al.\ 2018, arXiv:1808.02959
\bibitem[Chang et al.(2014)]{Chang14}Chang, P., Broderick, A. E., Pfrommer, C., Puchwein, E., Lamberts, A., \& Shalaby, M.\ 2014, ApJ, 797, 110
\bibitem[de Naurois(2015)]{Naurois15}de Naurois, M.\ 2015, Proceedings of the 
34$^\text{th}$ International Cosmic Ray Conference, Netherlands
\bibitem[Elyiv et al.(2009)]{Elyiv09}Elyiv, A., Neronov, A., \& Semikoz, D. V.\ 2009, Phys. Rev. D, 80, 023010
\bibitem[Frank et al.(2013)]{Frank13}Frank, M. R., de O{\~ n}a-Wilhelmi, E., \& Aharonian, F. A.\ 2013, Front. Phys., 8(6), 714
\bibitem[Funk(2015)]{Funk15}Funk, S.\ 2015, Annu. Rev. Nucl. Part. Sci., 65, 245
\bibitem[Gould \& Schr\'eder(1966)]{Gould66} Gould, R. J., \& Schr\'eder, S.\ 1966, PhRvL, 16, 748
\bibitem[Huba(2016)]{Huba16}Huba, J. D.\ 2016, NRL Plasma Formulary
\bibitem[Klimontovich(1982)]{Klimontovich82}Klimontovich, Y. L.\ 1982, Kinetic Theory of Nonideal Gases and Nonideal Plasmas (Oxford: Pergamon)
\bibitem[Miniati \& Elyiv(2013)]{Miniati13}Miniati, F., \& Elyiv, A.\ 2013, ApJ, 770, 54
\bibitem[Neronov \& Semikoz(2009)]{Neronov09}Neronov, A., \& Semikoz, D. V.\ 2009, Phys. Rev. D, 80, 123012
\bibitem[Neronov \& Vovk(2010)]{Neronov10}Neronov, A., \& Vovk, I.\ 2010, Science, 328, 73
\bibitem[Papadopoulos(1975)]{Papadopoulos75}Papadopoulos, K.\ 1975, Phys. Fluids, 18, 1769
\bibitem[Puchwein et al.(2012)]{Puchwein12}Puchwein, E., Pfrommer, C., Springel, V., Broderick, A. E., \& Chang, P.\ 2012, MNRAS, 423, 149
\bibitem[Rafighi et al.(2017)]{Rafighi17}Rafighi, I., Vafin, S., Pohl, M., \& Niemiec, J.\ 2017, A\&A, 607, A112
\bibitem[Schlickeiser et al.(2002)]{RS02}Schlickeiser, R., Vainio, R., B{\"o}ttcher, M., et al.\ 2002, A\&A, 393, 69
\bibitem[Schlickeiser et al.(2012a)]{RS12a}Schlickeiser, R., Elyiv, A., Ibscher, D., \& Miniati, F.\ 2012a, ApJ, 758, 101
\bibitem[Schlickeiser et al.(2012b)]{RS12}Schlickeiser, R., Ibscher, D., \& Supsar, M.\ 2012b, ApJ, 758, 102
\bibitem[Schlickeiser et al.(2013)]{RS13}Schlickeiser, R., Krakau, S., \& Supsar, M.\ 2013, ApJ, 777, 49
\bibitem[Shalaby et al.(2018)]{Shalaby18}Shalaby, M., Broderick, A. E., Chang, P., et al.\ 2018, ApJ, 859, 45
\bibitem[Sironi \& Giannios(2014)]{Sironi14}Sironi, L., \& Giannios, D.\ 2014, ApJ, 787, 49
\bibitem[Tavecchio et al. (2010)]{Tavecchio10} Tavecchio, F., Ghisellini, G., Foschini, L., et al. 2010, MNRAS, 406, L70
\bibitem[Tavecchio et al. (2011)]{Tavecchio11} Tavecchio, F., Ghisellini, G., Bonnoli, G., \& Foschini, L. 2011, MNRAS, 414, 3566
\bibitem[Taylor et al.(2011)]{Taylor11}Taylor, A. M., Vovk, I., \& Neronov, A.\ 2011, A\&A, 529, A144
\bibitem[Vafin et al.(2018)]{Vafin18}Vafin, S., Rafighi, I., \& Pohl, M.\ 2018, ApJ, 857, 43
\bibitem[Zakharov(1972)]{Zakharov72}Zakharov, V. E.\ 1972, Sov. Phys. JETP, 35, 908

\end{thebibliography}
\end{document}